\documentclass[12pt]{iopart}

\usepackage{iopams}
\usepackage{graphicx}
\begin{document}

\title[SPIRAL]{Nuclear structure and reaction studies at  SPIRAL}

\author{A.~Navin, F.~de Oliveira Santos, P.~Roussel-Chomaz\footnote {Present address: CEA-Saclay, DSM/DIR, F-91191
Gif sur Yvette Cedex, France}, and O.~Sorlin}

\address{GANIL, CEA/DSM-CNRS/IN2P3, Boulevard Henri Becquerel, F-14076 Caen Cedex~5, France}
\ead{navin@ganil.fr}
\ead{francois.oliveira@ganil.fr}
\ead{patrica.roussel-chomaz@cea.fr}
\ead{sorlin@ganil.fr}
\begin{abstract}
The  SPIRAL facility  at  GANIL, operational  since  2001, is
described briefly.  The diverse physics program using the re-accelerated (1.2 to 25
MeV/u)  beams ranging  from  He to  Kr  and  the  instrumentation
specially  developed for  their exploitation  are presented.  Results of
these studies,  using both  direct and compound  processes, addressing
various questions related  to the  existence of  exotic states  of nuclear matter,  evolution  of new
``magic numbers'',  tunnelling
of   exotic   nuclei,  neutron   correlations,   exotic  pathways   in
astrophysical  sites   and  characterization  of   the  continuum  are
discussed. The future prospects for the facility and the path towards SPIRAL2,
a next generation ISOL facility, are also briefly presented.

\end{abstract}

\maketitle
\section{Introduction: GANIL and the SPIRAL facility}
The GANIL facility  completed 25 years of operation in 2008.  In the
early years, the experimental  programs focused mainly on  the use
of intense  stable beams and  Radioactive Ion Beams  (RIB) produced
using projectile fragmentation. The  LISE   spectrometer~\cite{Anne}
and the SISSI device~\cite{SISSI} were used to produce RIB using the
``in-flight" technique.  The genesis  of the SPIRAL (Systeme de
Production  d'Ions Radioactifs Acc\'el\'er\'es en Ligne)  facility
was the need to  complement  the  available  facilities  at  GANIL,
and  open  new experimental  avenues in  Europe with  re-accelerated
radioactive ion beams.  The SPIRAL project  was jointly funded by
IN$^2$P$^3$/CNRS, DSM/CEA and  the Regional Council  of
Basse-Normandie in December 1993. A combination  of the  ISOL
method~\cite{Kofo} (Isotope  Separation On Line) and the
post-acceleration of these secondary  beams to energies around 25
MeV/u followed  in the foot steps of the  facility at Louvain la
Neuve~\cite{Proc10}. The  originality  of SPIRAL lies  in the
exploitation of fragmentation of the broad range of energetic heavy
ions available at GANIL,  for  the production  of  the  ISOL
secondary beams.   Such an approach, where the  projectile rather
than the target is varied to produce different radioactive species
differs  from the use of proton (or  light-ion) beams used  at the
other ISOL facilities~\cite{Spiralorg}. A large collaborative effort
between laboratories in France and Europe was established in order
to develop and construct the  instrumentation required to exploit
these beams. The relatively high energies of  the re-accelerated
beams, in  conjunction with  the dedicated  instrumentation has
allowed, as will be seen, to address challenging problems  in the
field to be addressed.

\begin{figure}
\includegraphics[width=\columnwidth] {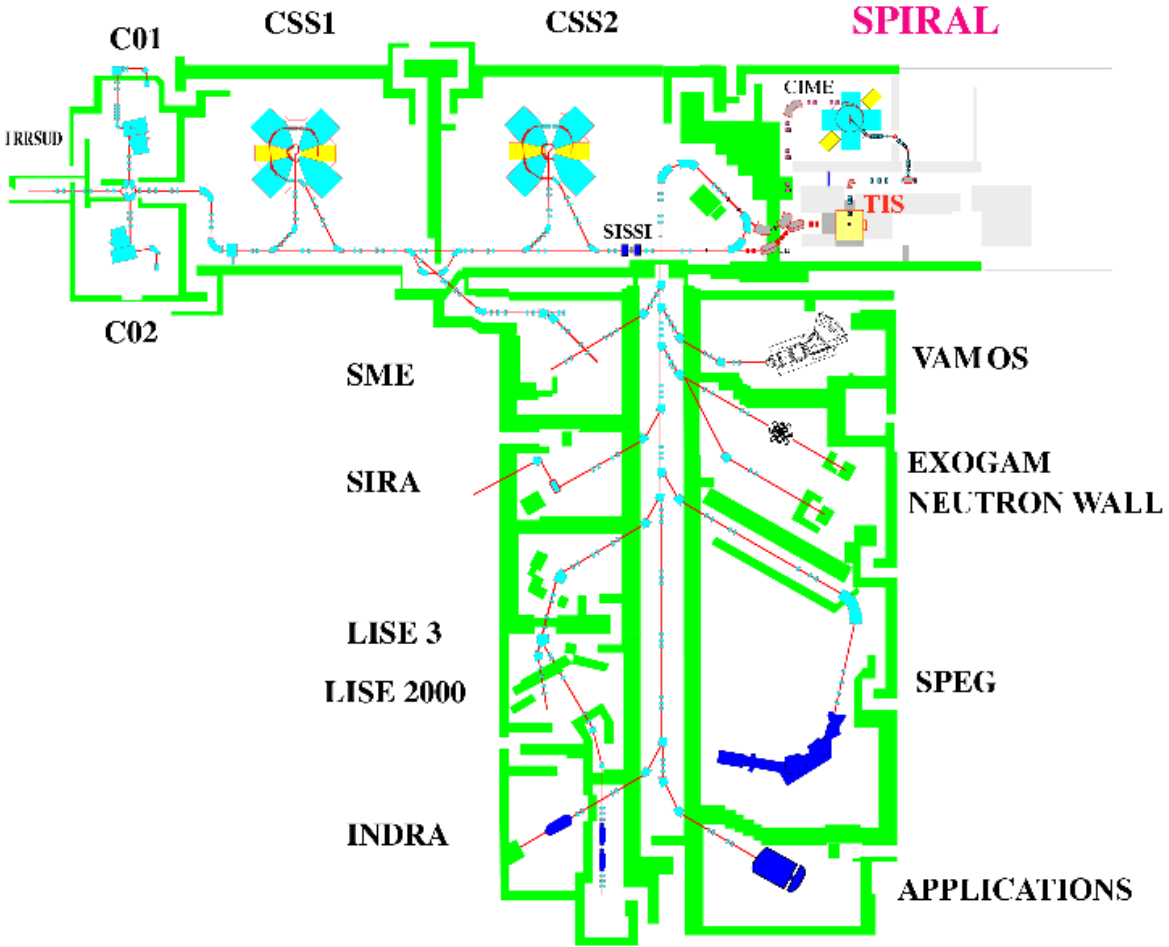}
\caption{(colour online) A  schematic view of the  GANIL facility.    C01 and  C02 are  the
  injector  cyclotrons.   CSS1  and  CSS2  are  the  sector separated
  cyclotrons used  to accelerate the  primary beam which  bombards the
  target  ion  source (TIS)  production  system  of  SPIRAL. The CIME
  cyclotron is used to post accelerate the radioactive ion beams which
  are further  momentum analyzed  before being  sent  to   any  of
 the  experimental   areas. (The SISSI device has recently  been decommissioned).
\label{spiralfig1}}
\end{figure}

\begin{figure}
\begin{center}
\includegraphics[width=\columnwidth] {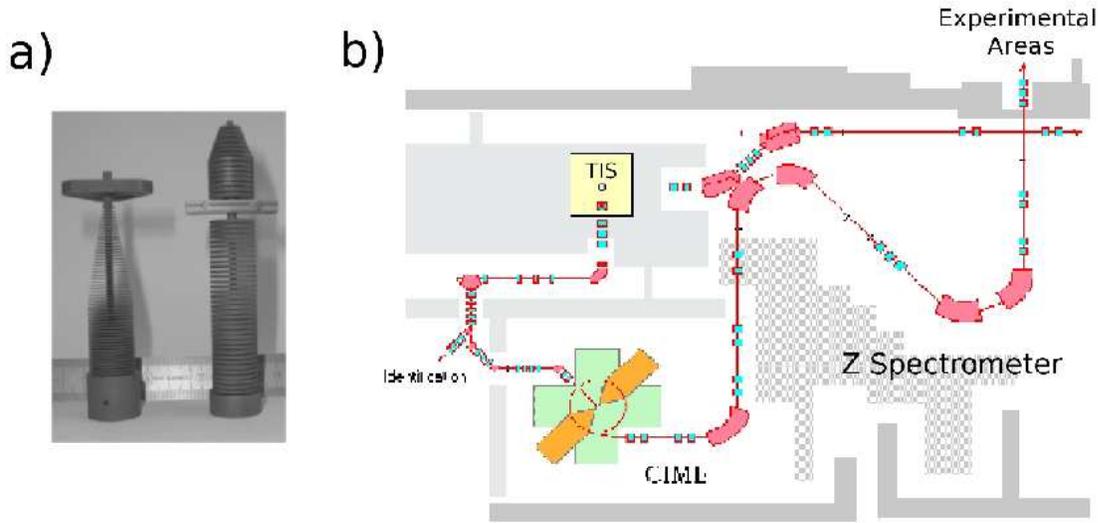}
\caption{(colour online) a) The SPIRAL graphite targets. b) Detailed schematic view of the SPIRAL facility.
\label{spiralfig2}}
\end{center}
\end{figure}

The number  of radioactive  atoms created by  the ISOL  method depends
both on  the primary beam  intensity and the  integrated fragmentation
cross section. The intensities of the primary beams available at GANIL,
($^{12}$C  to  $^{238}$U) vary  from  $6  \times  10^9$ to  $2  \times
10^{13}$~pps. These  high
intensity heavy-ion  beams accelerated  by the CSS1  and CSS2
cyclotrons (with a maximum  energy of 95 MeV/u), shown in Fig.~\ref{spiralfig1}, bombard a thick graphite production
  target   (Fig.~\ref{spiralfig2}).  The Target-Ion Source system (TIS)
is located underground  inside   a  well shielded area.
The  target,  where all  the  reaction
products are  stopped, is   heated mainly  by the  primary beam up  to a
temperature  of $2000^{\circ}$C. The conical shape  of the target
permits an optimum distribution  of the energy deposited by the beam,
especially in  the region  of the Bragg  peak. In cases where the primary beam
power is relatively low, a supplementary  extra ohmic heating  can be added through the axis  of the
target to maintain the diffusion of the  atoms produced.
After production  and diffusion, the  radioactive atoms effuse  to the
ion source  through a cold transfer  tube.  Most of
the non-gaseous elements are adsorbed here leading to a chemical selection.
The  atoms then  enter the 10 GHz  permanent-magnet ECRIS  (Electron
Cyclotron Resonance Ion Source) Nanogan-3, where they are ionized and
extracted~\cite{Nanogan}.    The radioactive  atoms are  ionized  up to  charge states
corresponding to q/m= 0.06 to  0.40.  After extraction from the ECRIS,
the  low-energy RIB  (acceleration voltage  from 7  kV to  34  kV) are
selected by a relatively    low     resolution    separator
($\Delta$m/m~=~4$\times$10$^{-3}$) and injected  into the CIME cyclotron
(Cyclotron  pour Ions  de Moyenne  Energie), constructed as a part of  the
SPIRAL project. In order to test  the properties of different target ion source configurations
under  experimental conditions  (limited  to a maximum of  400W  beam power)  a
separator  SIRa  (S\'eparateur  d'Ions RAdiactifs) was  constructed  in
1993.  A  detailed layout of
the SPIRAL facility and typical SPIRAL targets are shown in Fig.~\ref{spiralfig2}.
CIME is a room temperature compact cyclotron having  K = 265 (B$\rho$ =
2.344 Tm), a magnetic field between 0.75  and 1.56 T and
an  ejection radius  of 1.5m.  The  beams can  be accelerated within an
energy range  of 1.2  to 25 MeV/u, depending on  their mass-to-charge
ratio.  The beam emittance at the injection is limited to 80$\pi$ mm-mrad and at
ejection to $\sim$16$\pi$ mm-mrad.  The  magnet of CIME
has  4 return yokes  made of  thick, stacked
slabs.  This structure was chosen for its compactness, high field
homogeneity  and  low  fabrication   cost.   The  RF
resonators are of a classical  type with a  cantilever dee and  a sliding
short  circuit   using  spring  contacts  around   the  stems.  The
accelerated  RIB are  selected  based on their magnetic  rigidity by  the
$Z$-shaped  spectrometer (Fig.~\ref{spiralfig2}b) and directed  to  the existing  experimental
areas. Specially designed gas  profilers and sensitive  Faraday cups  are used
to monitor the  profile and intensity  of the  beams over  a wide  dynamic range
(between  10$^3$  and  10$^7$~pps). CIME itself acts as a  mass separator with a  resolving power
greater than 1/2500.  Additional purification can be achieved using a
stripper foil to select ions with different masses but having the same mass-to-charge ratio.
\begin{figure}
\begin{center}
\includegraphics[width=\columnwidth] {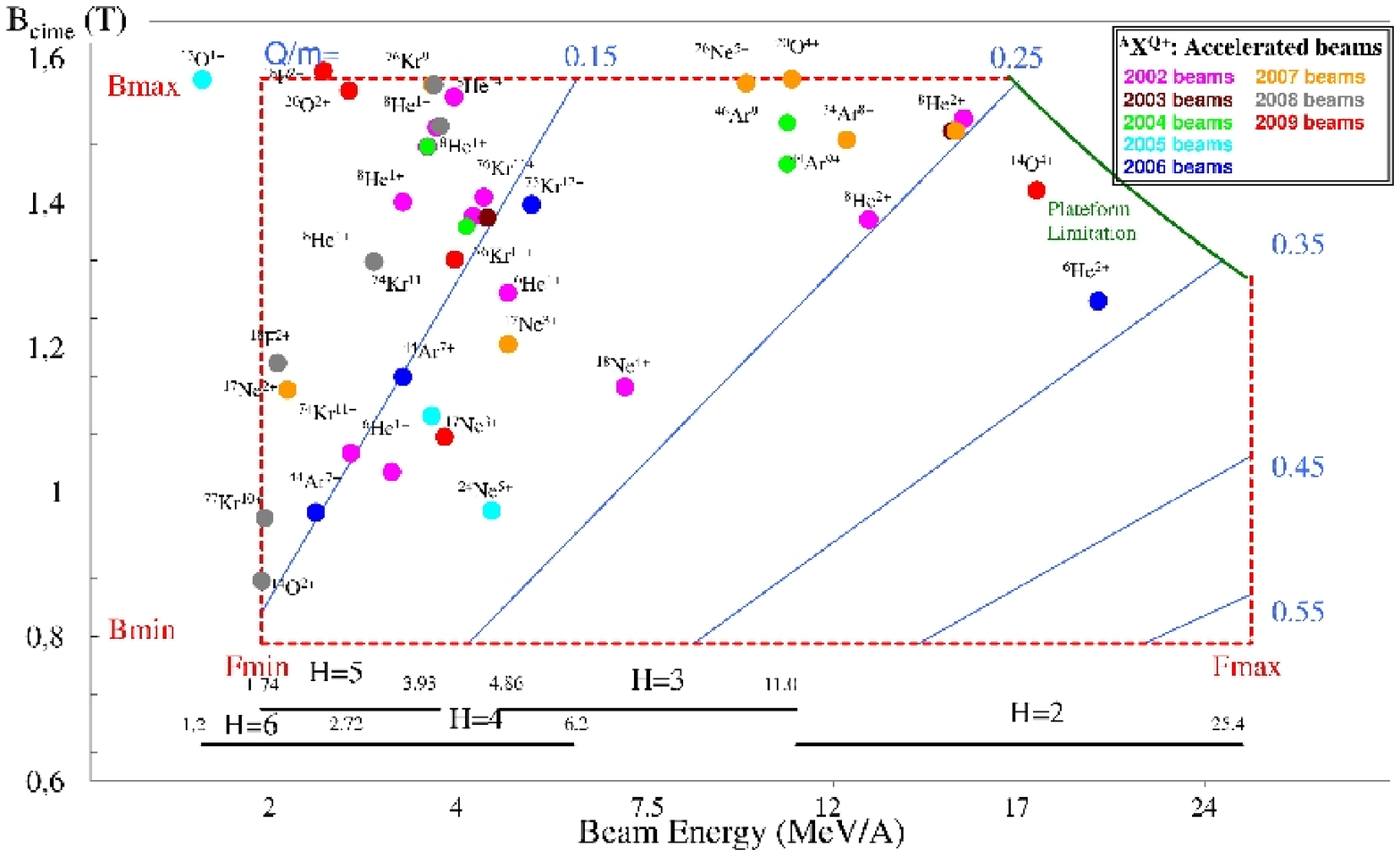}
\caption{(colour online) The operating  characteristics of CIME including the  radioactive ion species
that have been accelerated.
\label{spiralop}}
\end{center}
\end{figure}
The  first SPIRAL radioactive  ion beams  were delivered  in September
2001.  The intensities and variety  of beams that are available can be
found at:  http://pro.ganil-spiral2.eu/users-guide/accelerators/spiral-beams.
The  operational characteristics  of CIME,
and  the  various  beams  and  energies that  have  been  accelerated at SPIRAL are shown in Fig.~\ref{spiralop}.
Because of the chemical selectivity  inherent to the TIS, SPIRAL beams
were  initially limited  to  noble gases.   More recently  RIB of
Oxygen, Nitrogen  and Fluorine  have also been produced exploiting the fact that these  elements can be produced
as molecules in the graphite target. Developments to  further extend the
available  range of  elements which can be re-accelerated are in progress.
Restrictions  arising owing to safety  regulations  previously prevented  the   continuous  use  of  the  target
assemblies, thus substantially limiting the full exploitation of this
facility. These safety restrictions were reviewed and relaxed in 2007.

Although a majority of the experimental
programs at SPIRAL use  post-accelerated ISOL beams  from the CIME,
experiments  are also performed  without post acceleration.
Precision measurements of  the $\beta$-$\nu$  angular correlation parameter in $^6$He to search for signatures of physics beyond
the standard electroweak model~\cite{Oscar}, and the measurement of the charge
radius of $^8$He~\cite{Mueller}, are among the noteworthy results obtained  using these low energy (30 keV) beams.

The present article will first describe briefly the experimental equipment specially
constructed to exploit the SPIRAL beams. These include,
the large acceptance spectrometer VAMOS,  the segmented gamma array EXOGAM, the charged  particle  detector arrays
MUST/MUST2, TIARA and the  active target, MAYA.
The various  physics programs  using re-accelerated beams are then described. These include: a) The exploitation of direct
reactions  to measure  and understand  the spectroscopy  of  bound and
unbound states in  nuclei far from stability;  b) Reaction and spectroscopic studies
around  the Coulomb  barrier;  c)  Selected topics   in  nuclear  astrophysics,   probed  using   resonant  elastic
scattering  and transfer  reactions.  Finally,  the evolution  of  the  facility  and  its evolution  towards  SPIRAL2  are  briefly
presented.

 \begin{figure}
\includegraphics[width=\columnwidth]{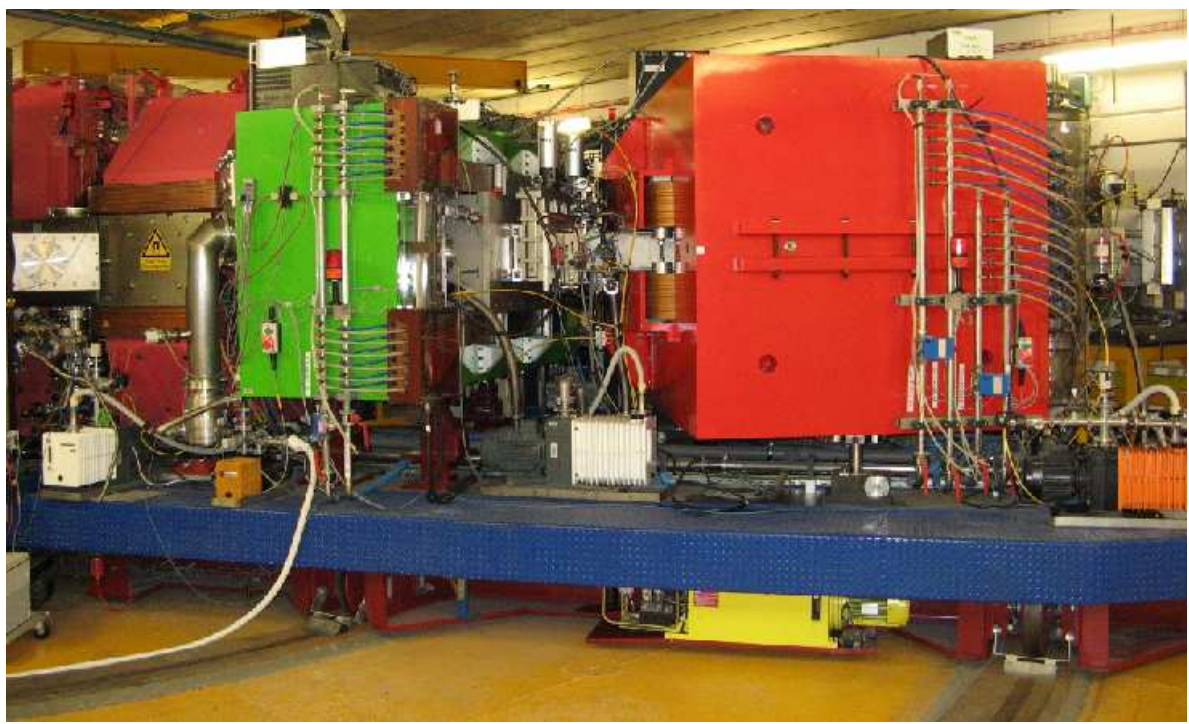}
\caption{(colour online) The VAMOS spectrometer at  GANIL showing (from left to right)
  the  two large  quadrupoles,  Wein filter,  Dipole  and part of the focal  plane
  detection chamber.}
\label{vamos}
\end{figure}

\section{Experimental equipment}
Intensities of re-accelerated radioactive ion beams, at best  four or five
orders of  magnitude lower than  typical intensities of  stable beams,
necessitate detection  devices  of   a  very  high  efficiency  and
granularity.   The 4$\pi$ $\gamma$-array  EXOGAM~\cite{Sim00}  and the
high   acceptance,   variable   mode   magnetic   spectrometer
VAMOS~\cite{VAMOS,Pul08} were  constructed with several European
collaborations.   Charged particle  detectors MUST~\cite{MUST}, MUST2~\cite{MUST2} and  the TIARA array \cite{TIARA} together  with the MAYA~\cite{Dem} active target have been the other major detectors developed
to exploit  the SPIRAL beams. Related developments  in electronics and
data  acquisition systems  have allowed  to couple  and  integrate the
above  independent  detector  systems   in  an  optimum  manner~\cite
{centrum}.

\subsection{VAMOS}

The  VAMOS  spectrometer  (VAriable  MOde  Spectrometer)~\cite{VAMOS,Pul08}
is a  large acceptance ($\sim$  60 msr) ray-tracing
spectrometer. The spectrometer has been primarily designed
for  a wide range of reactions
such as  elastic and  inelastic scattering, transfer,  deep inelastic,
fission and  fusion evaporation reactions at energies around the Coulomb barrier.
The nature  and kinematics (conventional
and inverse  kinematics) of  these diverse reactions with stable and
radioactive ion  beams demand
a high  performance  device  that detects and identifies  reaction
products  over a  wide range of energies and angles.
To meet these requirements, the  VAMOS spectrometer
combines  different modes  of operation in  a  single  device.
In the momentum dispersive mode of  operation, the  spectrometer
selects and  separates the  reaction
products  at the  focal plane  according to  their momentum  to charge
(p/q)  ratio  and  their  unique  identification is  achieved  via  an
event-by-event  reconstruction of the ion  trajectory in  the magnetic
field.  This  eliminates the need  for a position  sensitive detector
(near the target) to reconstruct  the scattering angle and thus allows
the  spectrometer  to also be  used  at  0$^\circ$.  When operating  as  a
velocity filter at 0$^\circ$,  VAMOS separates the reaction
products  from the direct  beam and  transports them  to the  focal plane.
The acceptance of the spectrometer has been studied by  simulating
the trajectory of the  particle using  an ion optical  code.  The versatile  detection system
allows for the  unique identification of  products (M and Z), ranging  from light  charged
particle to  fission fragments arising  in collisions induced  by both
stable  and  radioactive  ion  beams  from  both  ISOL  and  in-flight
fragmentation at  GANIL~\cite{SB08}. The detectors at  the focal plane
consist of large area  drift chambers, secondary electron detectors, a
segmented   ionization  chamber, and  Silicon   and  Plastic   detectors.   The
spectrometer  has  a maximum  nominal  rigidity  of  1.8 T-m,  B$\rho$
acceptance of  ~$\pm10\%$, an angular opening of  ~$\pm7^\circ$, and a
flight path  of $\sim$8 m.  The  spectrometer can be  moved around the
target to a maximum angle of  60$^\circ$.  VAMOS is operated in conjunction with a range
of  detector  arrays,  including EXOGAM,  MUST2, TIARA,  INDRA.  The VAMOS detection system is  presently been upgraded to
increase the  momentum acceptance of the spectrometer. Recently the
spectrometer was  successfully operated in a gas  filled mode to
increase  its  versatility, especially  for  experiments requiring  the
measurement of evaporation residues at 0$^\circ$ \cite{gasfilled}.

\begin{figure}
\begin{center}
\includegraphics[width=0.6\columnwidth]{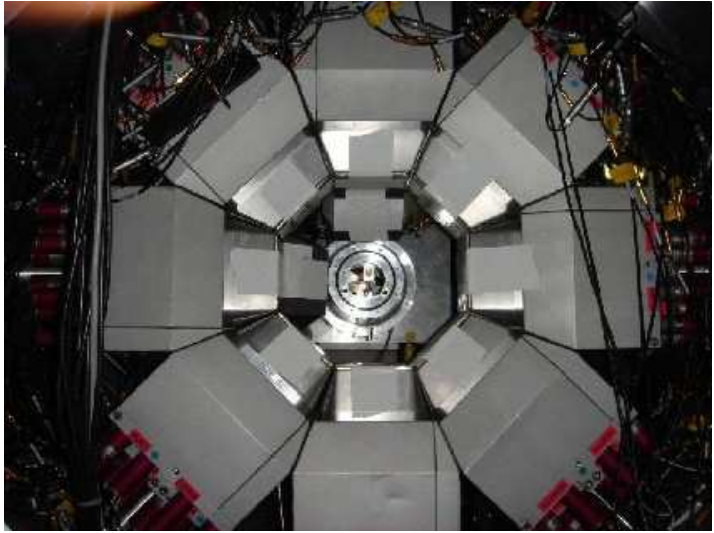}
\caption{(colour online) The EXOGAM array in a closed packed geometry.}
\label{figexogam}
\end{center}
\end{figure}

\subsection{EXOGAM}
The EXOGAM~\cite{Sim00}  array is composed of up to 16  segmented
Compton suppressed Ge clover  detectors. It was designed  to  comply
with the  special requirements imposed by  the use  of low intensity
radioactive ion  beams. Presently the array uses a maximum of 12
detectors. In particular, it was constructed to be a compact, highly
efficient and segmented device, most suitable for low multiplicity
events. The design also took into account the compatibility with
different experimental  setups  and   ancillary detectors.   The
16-fold segmentation of  each clover detector and the high  photo
peak efficiency make it ideal for the  search for characteristic
$\gamma$ rays produced in rare events.  Photo  peak efficiencies at
$\gamma$-ray  energies around 1~MeV are  $\sim$ 1 $\%$ per  detector
when placed  in a  closed packed structure (Fig.~\ref{figexogam}).
The high  granularity allows correction of the Doppler effect  for
the gamma  rays  emitted from fast moving products (typically $\sim$
20  keV for a 1  MeV gamma ray emitted at $\beta\sim$  0.1).   The
EXOGAM  detector  array has  been mainly used in two configurations,
one which maximizes the efficiency for the EXOGAM array
(Fig.~\ref{figexogam}) and another where  the peak-to-total ratio is
enhanced  with the  use of  the full Compton suppression to increase
sensitivity. An improvement in the position resolution, presently
limited by the  size of the electrical segment of the crystals was
obtained using a pulse shape analysis technique~\cite{Unsworth09}.
EXOGAM uses the VXI  bus standard developed  for the  EUROGAM and
EUROBALL  arrays to operate at high rates~\cite{exowebsite}. To
further improve the performance of this detector array, the signal
processing is  being upgraded from analog to digital processing.
This array is  operated in a stand  alone mode and  also in
conjunction with other detectors with  both stable beams and RIB
(produced in fragmentation reactions or through ISOL methods). It is
most often used with VAMOS but  has also been used in conjunction
with the LISE~\cite{Anne} and SPEG~\cite{SPEG} spectrometers.

\subsection{Tools for Direct reactions: MUST, MUST2, TIARA and  MAYA}
\begin{figure}
\begin{center}
\includegraphics[width=0.8\columnwidth]{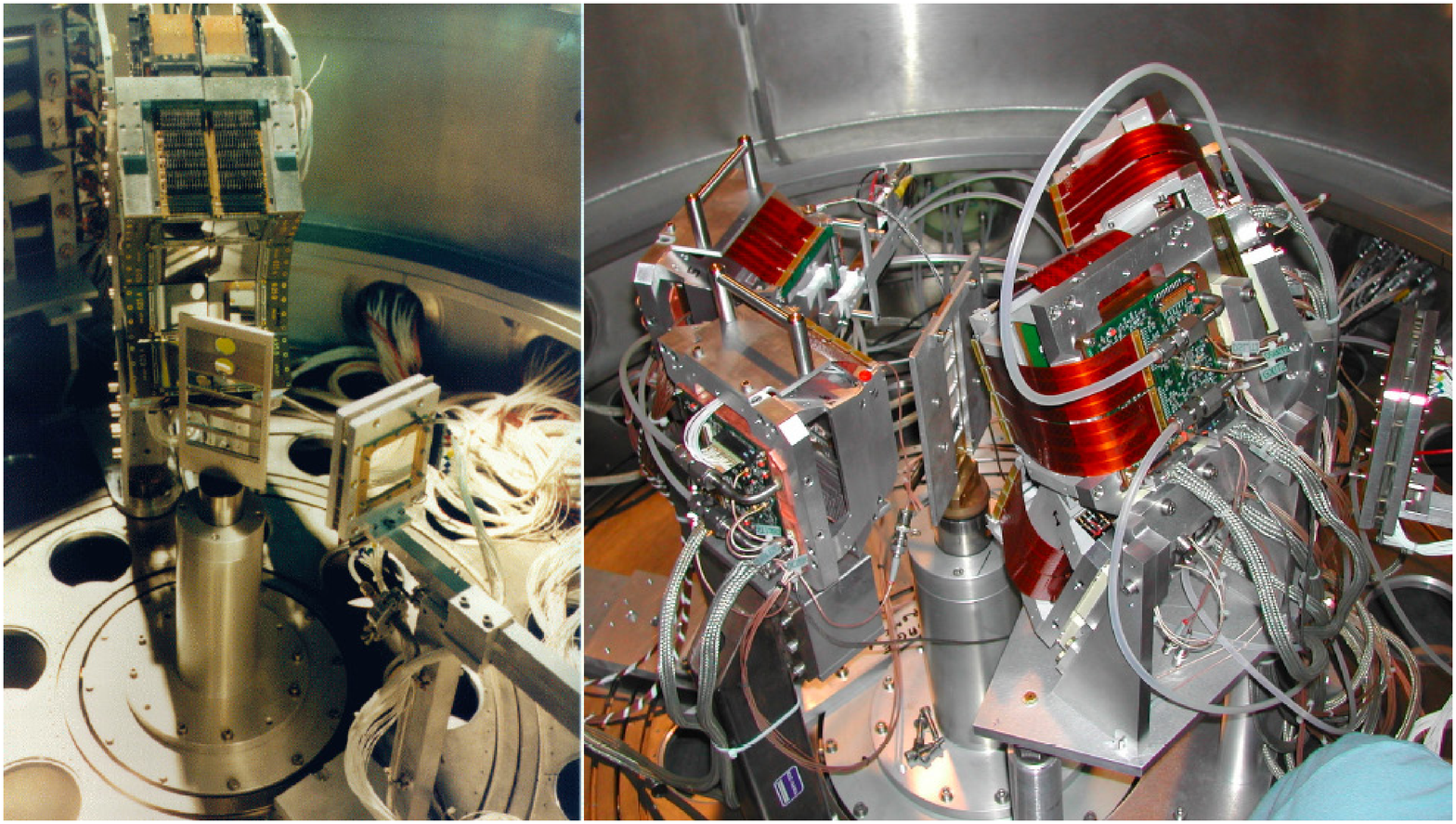}
\includegraphics[width=0.8\columnwidth]{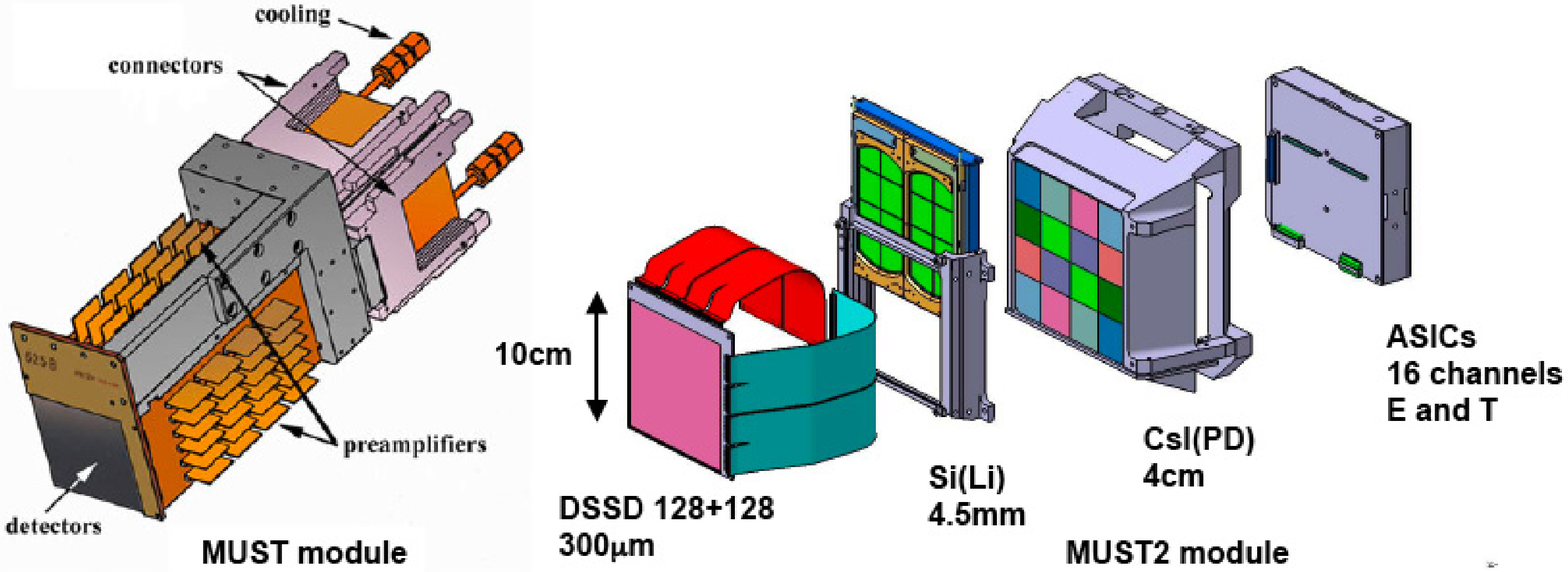}
\caption{\label{MUSTpict} (colour online) Top panel: The MUST (left) and MUST2 (right) detector
  arrays. Bottom panel: Schematic representations of single modules of  the MUST  and MUST2
  detector arrays.}
\end{center}
\end{figure}

New charged particle detector arrays with high efficiency and granularity
have been developed to exploit direct reaction studies,
taking   into   account   the   experimental   conditions  imposed by
re-accelerated  RIB.  The  experiments  are  performed  in  inverse
kinematics, with  the target nucleus playing  the role of the light beam in
a direct reaction in normal kinematics. The target may either be a thin foil of
CH$_2$ or  CD$_2$ in the case  of proton or deuteron  scattering, or a
cryogenic target of hydrogen or  helium.  As the nuclei under study are often
loosely bound, many states of  interest are located above the particle emission
threshold and  therefore decay in  flight, requiring, in  principle, the
complete kinematic coincidence detection of  the ejectile and the decaying
particles.  The alternative  method, used for direct reaction studies  with SPIRAL  beams,
is  the   missing   mass   method,  where   the
characteristics  of the  nucleus  of interest  are  obtained from  the
kinematical properties  of the recoiling light particle  in the two-body
reaction.  The detectors developed include the MUST, MUST2  and TIARA
solid state detector arrays and the MAYA active target.  Suitable beam
tracking detectors were also developed in order to improve the angular
definition  of the  incoming projectiles  on an  event-by-event basis~\cite{CATS}.

\subsubsection{MUST and MUST2 \\}

The MUST \cite{MUST} and MUST2 \cite{MUST2} detector arrays consist
of 8 telescopes composed of silicon strip, Si(Li) and CsI detectors.
These are used for  the  identification  of  the recoiling  light
charged  particles through time-of-flight,  energy loss and residual
energy measurements. The scattering  angle is determined  from the
horizontal  and vertical position   measurements   obtained  from
the   segmented  Si   strip detectors.  The  MUST  array   was
recently  superseded  by  the  MUST2 array. The  latter has a better
angular coverage and  resolution, and the smaller volume of the
array allows for coincidence experiments with gamma-detectors. Going
from MUST to MUST2, the size of the telescopes was  increased from
6x6  cm$^2$ to  10x10 cm$^2$,  the strip  size was decreased from
1mm down to 0.67 mm and finally the front-end logic and analog
electronics  was replaced  by a very compact  micro-electronics
technology (MATE ASIC chips~\cite{Asic}).  The 500 $\mu$m
double-sided strip Si  detectors of  both arrays have  an energy
resolution  of the order of 40-50 keV, time-of-flight resolution
around 1 ns, and an energy threshold of $\sim$~150 keV for protons.
The full energy range covered by the  MUST   (MUST2)  array
corresponds   to  70  MeV  (90   MeV)  for protons. These arrays
were constructed  in a modular manner in order to cover the angular
range of  interest with a relatively small number of detectors:
around  90 $\deg$  in the centre-of-mass  (c.m.)  for  elastic and
inelastic scattering; at forward angles for pick-up  reactions such
as  (p,d), (d,t),  or (d,$^3$He);  and at backward angles for
stripping reactions such as  (d,p).  Figure \ref{MUSTpict} shows
images  of the  MUST and MUST2  arrays, together  with  schematic
representations showing the various components of each module for
both arrays.

\begin{figure}
\includegraphics[width=0.8\columnwidth]{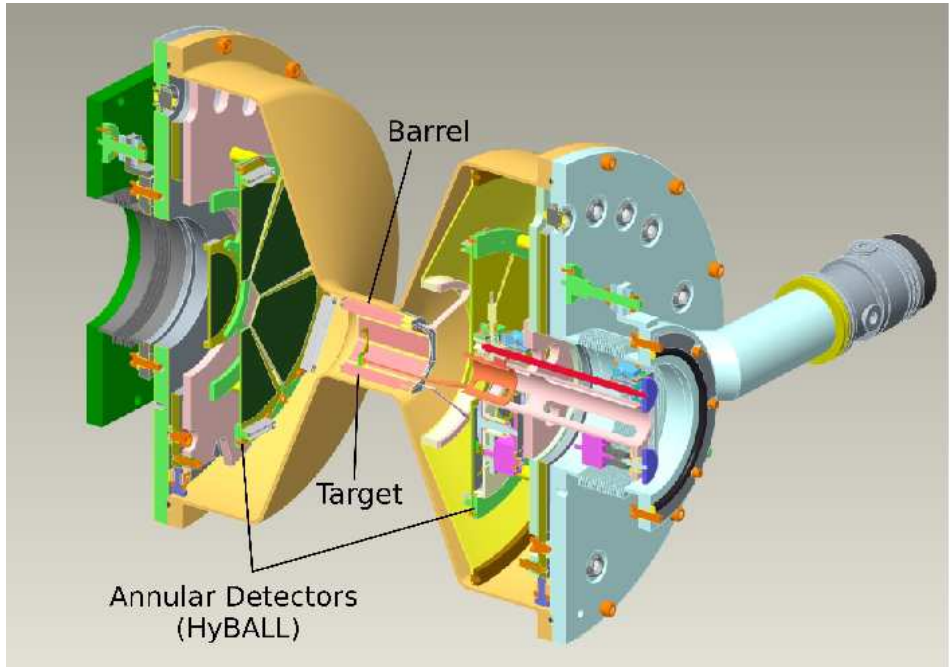}
\caption{ (colour online)  A schematic representation of the TIARA detector array.}
\label{TIARA}
\end{figure}
\subsubsection{TIARA \\}

TIARA  is  another   compact position  sensitive  Si  array,
designed to study direct reactions induced by radioactive ion beams in
inverse kinematics. While MUST and  MUST2 have been mainly designed to
be able  to perform inclusive  measurements of light  particles, TIARA
was  specifically conceived to  be used  in coincidence  with $\gamma$
rays detected  with  EXOGAM and the heavy  ejectile in the VAMOS
spectrometer.  It was constructed within a U.K led collaboration~\cite{TIARA}.   Figure~\ref{TIARA} shows  a
schematic representation of the  array.  It  consists of 8
resistive strip  Si detectors forming  an octagonal barrel  around the
target covering angles between 36$^\circ$  and 143$^\circ$.  Each detector has
an active area of about 95mm long and  22.5mm wide, with a thickness of
400$\mu$m.   The junction  side  facing  the target  is  divided into 4
longitudinal resistive strips, each with a width of 5.65mm. In  total the 32
strips provide  azimuthal angles in bins of  approximately 9.5 degrees
each.  In  addition, a set  of double-sided strip Si  annular detectors
can be  positioned at each end of  the barrel detector, complementing the
angular range of the barrel.  The TIARA reaction chamber has the shape
of a  longitudinal diabolo,  with a central  cylinder section  of 85mm
outer diameter  housing the barrel and two  500mm diameter cylindrical
sections at each end housing  the annular detectors.  This geometry  maximizes the gamma detection efficiency around the target.

 \begin{figure}
\begin{center}
\includegraphics[height=10cm]{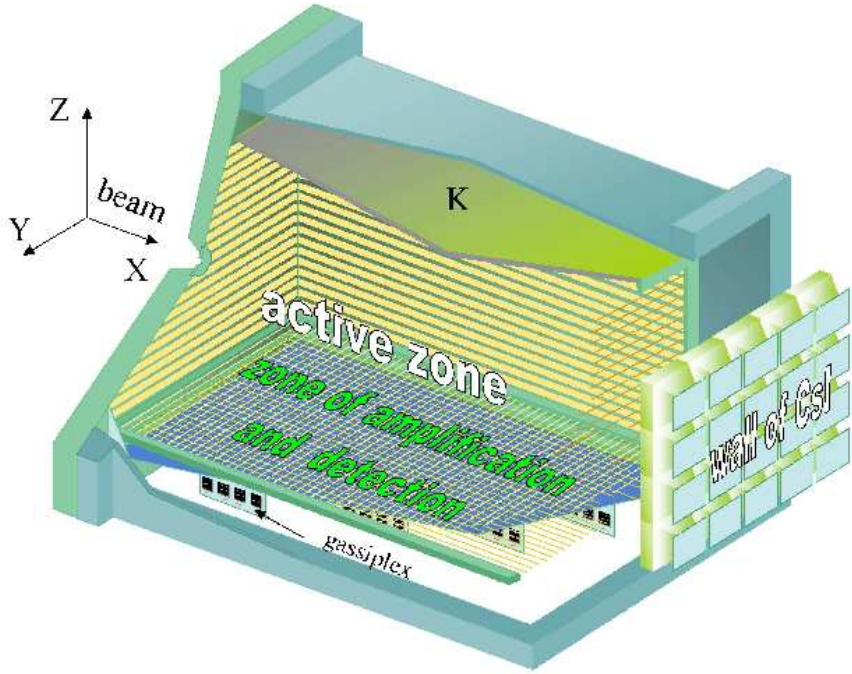}
\caption{\label{MAYA} (colour online)  Schematic view of MAYA active target including the forward angle CsI wall.}
\end{center}
\end{figure}

\subsubsection {MAYA: An active target\\}
An alternative method of  studying direct reactions consists of using
an active  target where the  gas atoms of  the target also
act  as the  detector. An active target in  principle has a  4$\pi$ angular
coverage, a very low detection threshold and can have a large effective
target  thickness without  a loss  of energy  resolution.  It  is thus
ideally  suited for  the study  of  reactions as  a three  dimensional
tracking of the reaction partners can be achieved.  In the MAYA active
target detector (Fig. \ref{MAYA}), developed at  GANIL~\cite{Dem}, the
3-dimensional  tracking of  the  incoming beam  and outgoing  reaction
partners  (ejectile  and  recoiling gas target)  is  attained  via  the
projection  of the  charge  created  on a  segmented  cathode for  the
horizontal  plane (32  rows  of 32  hexagonal  pads having  a size  of
$\approx$ 0.8  cm each), whereas  the vertical coordinate  is obtained
from the drift time  measurement on each individual amplification wire
with respect to the beam pulse. The induced charge on each cathode pad
is measured  using the GASSIPLEX chips~\cite{GASSI},  while the arrival
time of the electrons is registered with standard analogue electronics
on each amplification wire (one wire for each row of pads). For particles stopped
within the gas-volume, the angular  and range resolution is determined by the position  resolution.
Together they  determine the energy, the charge  and mass  resolution.  The  position resolution  is limited  to  about 1  mm
($\sigma$)  owing  to the  combined  effects of  the  pad  size and  the
digitization by  the amplifying wires.  For particles not  stopped in
the  gas volume, an array of  Si-CsI detectors,  configured differently for each
experiment cover the forward angles.  The identification  is obtained in this case from the
$\Delta$E-E  measurement, with energy  loss ($\Delta$E)  measured in  the last
section of the MAYA gas volume or in the Si detectors, and the residual
energy (E) in the Silicon or CsI detectors.  The detector has also been
operated using different gases, including He and CF$_4$.

\section{ Direct reaction studies in inverse kinematics to probe bound and unbound states
\label{transfer}}

The role  of the nucleon-nucleon (NN) interactions  in modifying shell
gaps in atomic nuclei is now well established \cite{Prog08}.  The most
salient feature  of these modifications appears  in the disappearance
of conventional ``magic numbers''  far from the valley of stability.
Transfer reactions such as (d,p) or (d,t) can be used to identify states
in nuclei  which have a large  overlap with pure  single-particle states.
They thus  provide  suitable inputs to the understanding of  nuclei far from stability, namely:
(i) the determination of the  in-medium (effective) NN interactions and
increasing the predictive power of structure calculations; (ii)  the deconvolution of
these  interactions  into central,  spin-orbit  and  tensor terms  and
identifying  the  role  of  each  of  these  in  the  shell
evolution;  (iii) the comparison of  the properties  of ``bare''  and in-medium
interactions to ultimately describe  the atomic nucleus from realistic
and not purely effective interactions; and (iv) the understanding of the role
of the continuum when approaching the drip line.

\begin{figure}
\begin{center}
\includegraphics[width=9cm]{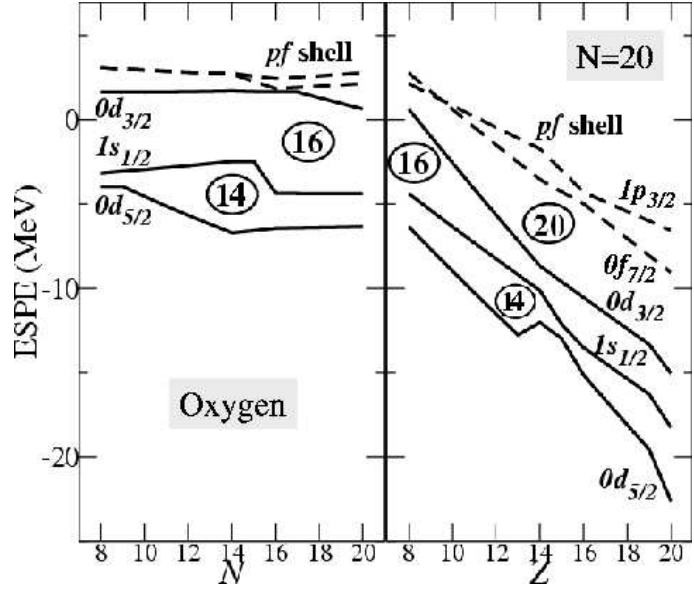}
\caption{Effective Single-Particle Energies (ESPE) of neutrons in the
  O isotopic chain   (left)  and  for  the  $N=20$   isotones  with  8$<$Z$<$20
  (right)  (Refs. \cite{Utsu99,Otsu02}). \label{espe}}
\end{center}
\end{figure}

The (d,p) reaction is used to populate valence neutron orbitals, whereas in the
(d,t) reaction  a neutron is  removed from  occupied  orbitals. The
angular  distributions of  the  measured protons (for  (d,p))  or tritons  (for
(d,t)) provide a signature of the angular momentum $\ell$ of the final
states populated,  while the  comparison of the  measured differential
cross  sections with  model  calculations (such as the Distorted  Wave Born Approximation or the Coupled Reaction
Channels formalisms)   quantitatively   determines  the  single-particle
strengths (spectroscopic factors).   Conventionally,
single-particle  states  have spectroscopic factors (SF)  values  close to  one.
Such studies as a function of proton number (for a fixed neutron
number) probe the change of  energies of neutron single-particle states arising from
the proton-neutron $V^{pn}$  interactions (right panel of Fig.  \ref{espe}).
Similarly, the evolution of neutron-single particle  energies as
a function of the neutron  number (for  a fixed $Z$)  can  be  used  to
determine  the  neutron-neutron  interactions $V^{nn}$ (left panel of Fig.\ref{espe}).
Examples of this evolution are illustrated in two regions of the chart
of  nuclides.

Experiments  carried  out  using  beams  of  $^{20}$O and
$^{24,26}$Ne to study the $N=14,16, 20$ shell  gaps are presented here.
This is  followed by the use of  $^{44,46}$Ar beams to probe
the $N=28$  shell gap and  the spin-orbit splitting. The analysis of the changes in
shell gaps employs principally a shell model approach. The measurements were made
at beam energies of  $\sim$10~MeV/u  where    direct  transfer
is expected  to be the dominant reaction  mechanism
and  permits the study of  excited states  up to $\sim$6~MeV.
Additionally, this also favours an angular momentum matching of around
$\ell =2$, which is ideal for the study of
the $sdpf$  shells.   The     TIARA~\cite{TIARA}    and    MUST~\cite{MUST}    or
MUST2~\cite{MUST2}   charged-particle    arrays   were   used   either
independently or  together to detect the light  charged particles. Tracking of
beam  particles, where necessary, was made using  the position sensitive  gas filled
detectors CATS~\cite{CATS}. The SPEG \cite{SPEG} and
VAMOS \cite{Pul08}  spectrometers were used  to  select and  identify the  heavy
transfer residues.  In some  cases, the EXOGAM  detectors were
used to achieve  a better final energy resolution. This also allowed
the spin of the states to be constrained using $\gamma$-decay  selection rules
(in addition  to the measured transfered angular momentum $\ell$).

\subsection{Addressing the $N=14, 16$ and $20$ (sub)shell gaps
\label{141620}}

The  variation of Effective  Single Particle  Energies (ESPE)  for the
$N=20$ isotones is shown as a function of the proton number $Z$ in the
right panel  of Fig.~\ref{espe}.  These energies are derived from the
shell-model  monopole   interactions~\cite{Utsu99,Otsu02}, constrained by
experimental data available up to 1999.  The remarkable
constancy of  the neutron $N=20$  shell gap between $Z=14$  and $Z=20$,
 as seen from  the figure, indicates the  doubly closed shell nature of  $^{34}_{14}$Si, $^{36}_{16}$S and
$^{40}$Ca.  For values of Z less than 14 the
$N=20$ gap  can be  seen to decrease.  Therefore, a neutron  from the
normally  occupied  $d_{3/2}$  shell  can  easily be promoted to  the  higher lying  $fp$
shells. Such correlations  across  the  N=20 shell  gap  lead to  the
existence  of deformed nuclei  in the  so-called ``island  of inversion''.
The disappearance of the $N=20$ shell gap is caused by the steep change of
the  neutron $d_{3/2}$  ESPE slope  as the  proton $d_{5/2}$  orbit is
filled.  This  is  a consequence  of  the largest  value of  the
corresponding  monopole  term  $V^{pn}_{d_{5/2} d_{3/2}}$ in  the  $sd$
shell. While the  $N=20$ shell gap reduces, a  new shell (or subshell)
gap  is expected  to appear  at $N=16$,  leading to a new  doubly magic
nucleus  $^{24}$O.   Despite being  smaller  in  magnitude, an  $N=14$
subshell  gap is  also present  for the  $Z=8-12$ nuclei.  However, as
shown in the  left panel of Fig.~\ref{espe}, this   subshell gap
does \emph{not}  exist at $N=8$ ($^{17}$O) but  appears when neutrons
are   added  to  the  $d_{5/2}$   orbital.   The   strongly  attractive
$V^{nn}_{d_{5/2} d_{5/2}}$ monopole  term increases the binding energy
of the $d_{5/2}$ orbit,  while the repulsive monopole $V^{nn}_{d_{5/2}
  s_{1/2}}$ reduces the binding  energy of the $s_{1/2}$ orbit leading
to  a $N=14$  gap between  the $s_{1/2}$  and $d_{5/2}$  orbits.  This
$N=14$ subshell  gap ($\sim$4~MeV in  $^{22}$O~\cite{Stan04}) leads to
$^{22}$O$_{14}$ behaving like a quasi-doubly  magic nucleus, and
accounts  for  the  spherical  nature of  $^{24}$Ne$_{14}$  while
$^{20}$Ne$_{10}$  is deformed.  Finally, the N=28  gap (between
$f_{7/2}$ and  $p_{3/2}$ orbitals),  which is  only $\sim$ 2~MeV in  $^{40}$Ca,
is  expected to be  further reduced at  lower $Z$
when  the $d_{5/2}$  orbit is  empty. These features are
addressed in Sects.~\ref{24Ne} and \ref{26Ne}.

\subsubsection{The $^{24}$Ne(d,p) reaction\label{24Ne} \\}

\begin{figure}[ht]
\begin{center}
\includegraphics[scale=0.7]{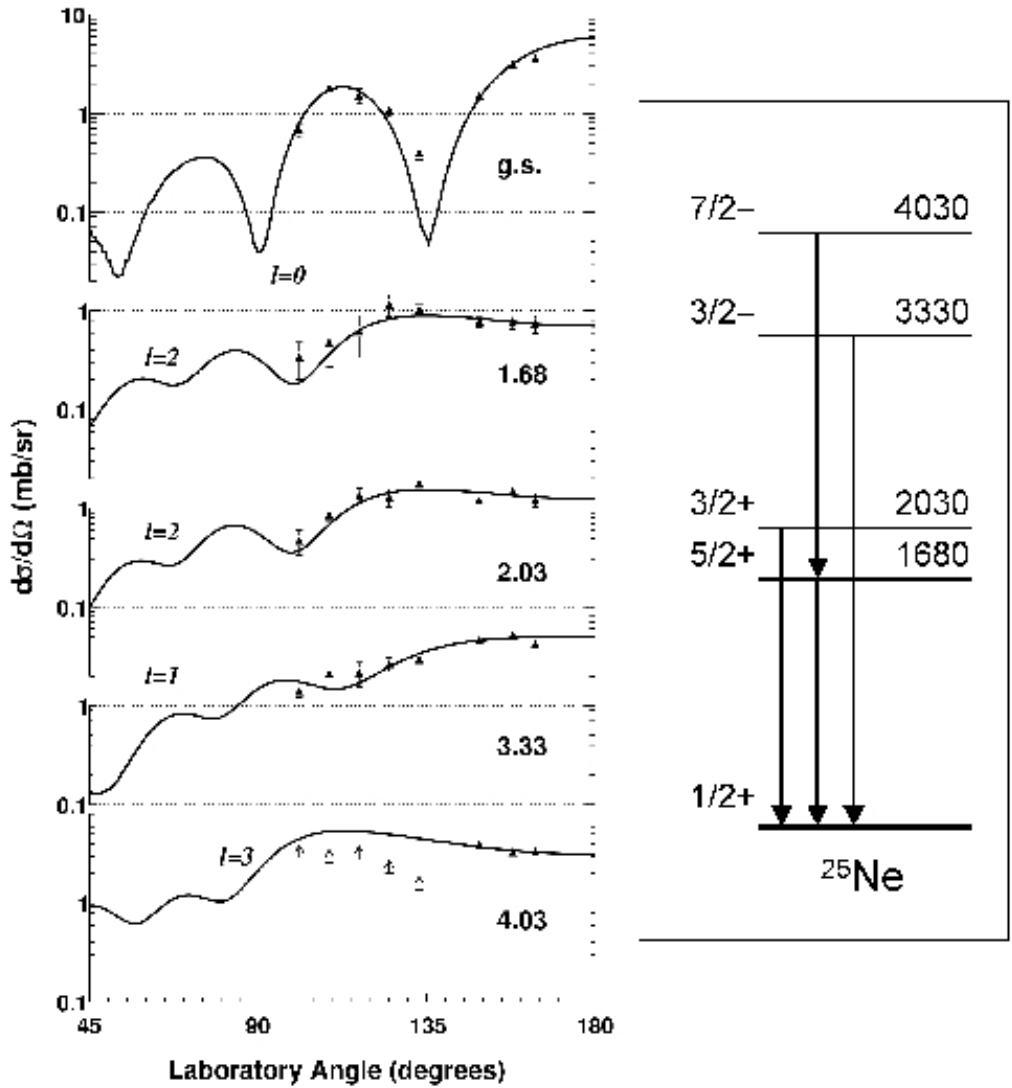}
  \caption{Left:  Proton angular  distributions for states populated in  the $^{24}$Ne(d,p)  reaction~\cite{Catf10}.
  The spectroscopic factors are noted for each state for the adopted $l$ transfers (solid lines from ADWA calculations).
  Right:
    Level       schemes deduced for $^{25}$Ne~\cite{Catf10,Terr04}.\label{24Ne_results}}
\end{center}
\end{figure}

Measurements  were performed  using a  10.5~MeV/u  $^{24}$Ne beam
(10$^5$ pps)  on a 1 mg/cm$^{2}$  CD$_2$ target.  The  charged particle array
TIARA  was used  to detect the  protons between  100$^\circ$ and
170$^\circ$  with  respect to  the  beam  direction.  It was  used  in
combination with  four segmented clover detectors of  the EXOGAM array
placed at 90$^\circ$  around the target (with a  photo peak efficiency
of  $\sim$8\%  at  1~MeV,  and   $\sim$50  keV  resolution  after  Doppler
correction).  The  heavy transfer  residues  ($^{25}$Ne) were  detected  and
identified  at  the  focal   plane  of  the  VAMOS  spectrometer.  The
high  detection efficiency  of the whole  system allowed
three-fold coincidence data ($\gamma$ + proton + $^{25}$Ne) to be acquired. The
measured angular  distributions for the  observed states are  shown in
the left panel of Fig.~\ref{24Ne_results}. The $\ell$ values as well
as the  spectroscopic factors of the corresponding states were  extracted from a  comparison with
Adiabatic Distorted  Wave  Approximation calculations (ADWA).  Combining $\gamma$-ray  branching  ratios,
angular  distributions and SF
values (with uncertainties of $\sim20\%$)  complete  spectroscopic  information  on $^{25}$Ne  has  been
obtained  and is shown in  Fig.~\ref{24Ne_results}~\cite{Catf10}).

Also shown in the figure is that the ground state of $^{25}$Ne has
the expected $\nu$$s_{1/2}$ configuration, arising from an $\ell$ =0
transfer with a large  SF of 0.8. Significant  SF's were also found
for the $3/2^-$ (0.75) and $7/2^-$ (0.73) states at 3.33 and 4.03
MeV respectively.  The  transfer to  the ``a priori'' fully occupied
$d_{5/2}$ orbit  is still possible, as  can be seen from the
measured SF value  of 0.15 for the $5/2^+$ hole state. A strictly
closed shell would have  given SF=1 for the $1/2^+$ state and 0 for
the $5/2^+$ state at 1.68 MeV.  Deviations from these values
indicate the possibility of generating neutron excitations across
the  $N=14$ gap.  The SF for  the $3/2^+$ state at 2.03 MeV is about
half  of that for a  pure $d_{3/2}$ single particle strength, the
remaining (according to Shell Model calculations) is fragmented
between several states at higher energies, each individually having
a small SF.  Except for the 3/2$^+$ state, the energies and SF of
the positive parity  states are in relatively good  agreement with
the shell  model calculations  using the  USD
interaction~\cite{USD}. The energy of  the 3/2$^+$ state  was used
as  a constraint to  adjust the monopole terms   involving  the
$d_{3/2}$  orbit   in  the   USD-A and USD-B
interactions~\cite{USDAB}.  The spin assignments  for  the $5/2^+$
and $3/2^+$  states were ascertained by the fact that only the hole
state $5/2^+$ is populated in  the one-neutron removal  from
$^{26}$Ne~\cite{Terr04}. The salient features  of the present work
can be summarized as follows : (1) the $N=14$ gap  is relatively
large in $^{25}$Ne (though smaller than that  in $^{23}$O by about 1
MeV); (2) the newly observed $3/2^+$ state  allows for a better
determination of the  monopole terms involving the  $d_{3/2}$ orbit,
thus providing for  a better understanding of  the evolution of the
$N=16$ gap; (3)  the $N=20$ gap formed between the $d_{3/2}$  and
intruder $fp$ states seems to be small (about 1.3~MeV); and (4) the
$N=28$  gap, that is 2 MeV in $^{40}$Ca, has greatly  decreased in
the Ne isotopes. This leads to an inversion between the $f_{7/2}$
and the $p_{3/2}$ orbits, as inferred from  the energies of the
$3/2^-$ and $7/2^-$ states.  Note that this inversion was predicted
by the ESPE of Fig.~\ref{espe}.

\subsubsection{The $^{26}$Ne(d,p) reaction \label{26Ne}\\}

\begin{figure}
\begin{center}
\includegraphics[scale=0.6]{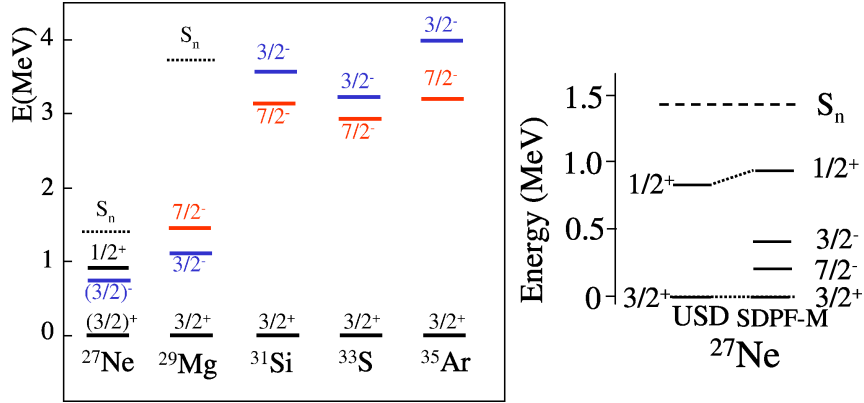}
\caption{(colour online) Left : Measured energies  of the  excited states  in the
  $N=17$  isotones. The  decrease in energy of  the $3/2^-$  and $7/2^-$
  intruder  states with  respect  to the  ground  state~\cite{Ober06} can be seen.
 Right : Shell model calculations using the SDPF-M
  interaction~\cite{Utsu99} for $^{27}$Ne. These calculations
  can account for  the presence of intruder states that are
  outside the valence space of the USD~\cite{USD} interaction.}
\label{26ne_results}
\end{center}
\end{figure}

This study provided the first evidence for the presence of an intruder
negative-parity state at very low energy in $^{27}$Ne. The energies of
the excited states in the $N=17$ isotones, showing the decrease of the
$3/2^-$ and $7/2^-$  intruder states with respect to  the ground state
are  shown in  Fig.  \ref{26ne_results}.  The  $^{26}$Ne(d,p)$^{27}$Ne
reaction~\cite{Ober06}  was  performed using   a  thick cryogenic  $D_2$
target.   The transfer  to  the bound excited states  was  obtained from  the
measurement of the Doppler-corrected $\gamma$-rays   in   the   EXOGAM
array. The  transfer-like  products  were detected  in  the  VAMOS
spectrometer.
The  cross  section for  the state  at 765~keV  leads to SF~=~0.6~$\pm$~ 0.2.
This is  consistent with  the  shell model
calculations  using SDPF-M  interaction  for a  negative parity  state
($3/2^-$ or $7/2^-$). A  smaller spectroscopic factor, SF~=~0.3~$\pm$~0.1,
was  found for  the  state  at  885~keV  for which  a  spin
assignment of $1/2^+$ was suggested based on its excitation energy and
spectroscopic  factor. Indeed, a  relatively small  SF for  the $1/2^+$
state  is expected,  as  the $\nu$$s_{1/2}$  orbit  is,  in principle,  fully
occupied  when  pair  scattering  to  the  upper  $d_{3/2}$  state  is
negligible.  As the protons  were  stopped in  a  thick target,  angular
distribution could not  be used to deduce the  spin assignments of the
observed states.  Complementary information on $^{27}$Ne from a knockout
reaction study    is   discussed   in   Ref.  \cite{Terr06}.   The  proton  angular
distributions  from the  $^{26}$Ne(d,p)$^{27}$Ne  reaction were recently  studied
using a SPIRAL beam with an intensity of $\sim$10$^3$~pps on a thin CD$_2$ target. The
TIARA  and EXOGAM  detector arrays  were  used to  detect protons  and
$\gamma$-rays respectively. Preliminary results from this work suggest that the 7/2$^-$
state is located 330$\pm$80~keV above the neutron-decay threshold (E$_x$=1.74$\pm$0.08~MeV)~\cite{Brown2010}.

\subsubsection{The $^{20}$O(d,t) and (d,p) reactions \label{20O}\\}

The  combined  study  of  $^{20}$O(d,t)  and  $^{20}$O(d,p)  reactions
provides  information on  the  single-particle  energies  of both  the
occupied  and valence  neutron  states  at $N=12$  in  the O  isotopic
chain. This can be compared to the known values at $N=8$ to derive the
evolution of the $N=14$ and  $N=16$ (sub)shell closures as the neutron
number is increased. Additionally, this work can  provide constraints
on the neutron-neutron interaction involved in the formation  of the
$N=14$ gap  and contribute to the understanding of the particle  instability of
$^{28}$O.

\begin{figure}
\begin{center}
\includegraphics[width=\columnwidth]{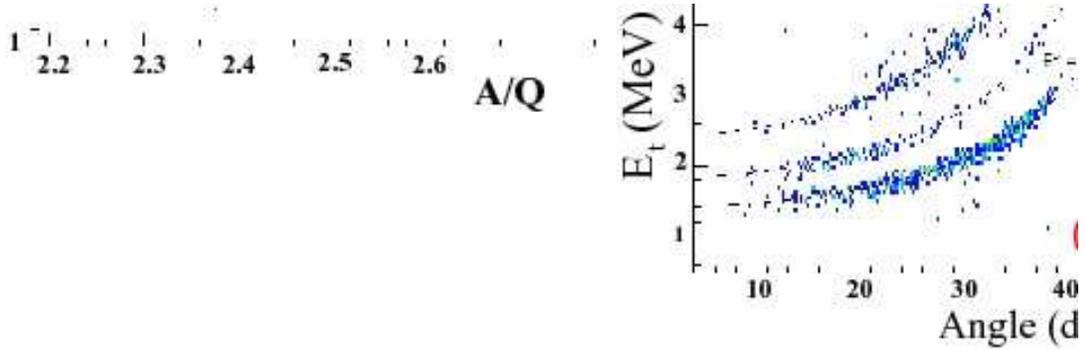}
\caption{(colour online) (a) Identification plot of energy loss versus A/Q obtained using VAMOS.
 The different reaction channels arising from
  the  $^{20}$O  transfer  reactions  can be clearly  identified.  A small
  contamination  of  $^{15}$N in  the  beam  can  also be  seen.
(b) Kinematic  plot of  the proton  energy as  a function  of
  angle  in the  laboratory  frame  obtained  by
  gating on $^{21}$O  in VAMOS. (c) Same as in  (b) showing the states
  in $^{19}$O produced by the (d,t) reaction.  These  preliminary  results are from
  Refs.~\cite{Bea,Ram08}} \label{20O_results}
\end{center}
\end{figure}

In this work, the TIARA and MUST2 detector arrays were placed around
a CD$_2$ target  to detect the  recoiling light particles between 36
to 169$^\circ$  and  8 to  36$^\circ$,  respectively. Additionally,
four EXOGAM clover detectors  were placed in a ``cube''  geometry
around the target, yielding a  photo peak efficiency of $\sim$8\%
at 1~MeV. A complete identification of all reaction channels was
achieved by an unambiguous detection of  the heavy products  in the
VAMOS  spectrometer, as  shown in Fig.~\ref{20O_results}a).
Detailed  analysis and  interpretation  are in  progress, however
some preliminary    results   from    Ref.~\cite{Ram08}   are shown
in Fig.~\ref{20O_results}.   Kinematic correlations  between the
energy and angle of the tritons (Fig.~\ref{20O_results}c) show the
energies of the states  produced in  the  (d,t)  reaction. The
lowest lying locus in the $^{20}$O(d,t) reaction corresponds to the
doublet formed by  the ground and an  excited state at 89~keV.
Gamma-ray  measurements indicate  that  the main  contribution  to
this  doublet arises  from a transfer  to   the  ground  state.
Other  transitions are  identified at $\sim$1400~keV and
$\sim$3200~keV.       As   can be seen in Fig.~\ref{20O_results}c,
the angular distributions of   the corresponding levels are quite
different. Angular momentum assignments $\ell =2,  0, 1$ have been
derived  for the ground + 89 keV doublet, 1400~keV and  3200~keV
states,  respectively.  The energies  of  the  observed states,
except for  the negative-parity states ($\ell=1$),  which are not in
the  valence space, are  found to be  in good agreement  with shell
model calculations  using the USD  interaction.  The evolution  of
the energy of the $\ell=1$ states  (from the occupied $p_{1/2}$
shell) may provide  information on  a possible  modification of  the
size  of the $N=8$   shell   gap  (between  the $\nu$$p_{1/2}$   and
$\nu$$d_{5/2}$ orbitals)  resulting  from the neutron-neutron
interactions. The values of the SF  for the  $\ell=0$ (1/2$^+$) and
$\ell=2$  (5/2$^+$  and  3/2$^+$  doublet)  states  also  provide
information   about   the   strength    of   $N=14$   shell   gap in
$^{20}$O.

Preliminary results on  the $^{20}$O(d,p) analysis are shown
in   Fig.\ref{20O_results}  (b).   Gating   on  $^{21}$O, as identified using VAMOS, two  clear
kinematical lines, which most likely characterize the transfer to $d_{5/2}$
and $s_{1/2}$ bound  orbitals, can be seen. A gate  on $^{20}$O, indicated a
broader kinematic line at a higher excitation energy (not shown). This
may corresponds to transfer to the neutron unbound $d_{3/2}$  state(s)  in
$^{21}$O.  These  results will provide further information on the
$N=14$ subshell gap (complementary  to the (d,t) data), as well as
the location  of the $d_{3/2}$ orbit and on the  size of the $N=16$
gap.  It could also provide further constraints on the neutron-neutron
interactions  involving  the unbound $d_{3/2}$
orbit, possibly providing a better insight  on the role of T=1 three-body
forces~\cite{Otsu10}.

\subsection{The evolution of the $N=28$ shell closure
\label{28}}

The ``magic number'' $28$, seen in the doubly closed shell nucleus $^{48}_{20}$Ca
was attributed to the effect  of the one body  spin-orbit (SO)  interaction.
Additionally, the $N=28$ gap is known to increase  by about 3~MeV
between $N=20$  ($^{40}$Ca) and  28 ($^{48}$Ca)   while
neutrons are  added to the $f_{7/2}$ orbit.   This behaviour arises  from the
attractive  nature  of  neutron-neutron monopole  $V_{f_{7/2}f_{7/2}}$
while the  $V_{f_{7/2}p_{3/2}}$ is repulsive~\cite{Prog08}.   In fact,
the mechanism which  forms the $N=28$ gap is the same  as that for the
$N=14$  gap shown in  the left  panel of  Fig.~\ref{espe}.  It  can be
viewed by replacing, in  this figure, the neutron ($d_{5/2}$, $s_{1/2}$)
orbits at $N=14$  by the ($f_{7/2}$, $p_{3/2}$) orbits  at $N=28$. The
evolution of  the $N=28$  shell gap below  $^{48}$Ca can be linked  to the
action of the $V^{pn}$  interaction   involving protons and neutrons
in the $sd$ and $fp$  shells, respectively~\cite{Prog08}.
The measurements of the $^{44}$Ar(d,p)$^{45}$Ar~\cite{45Ar}
and  $^{46}$Ar(d,p)$^{47}$Ar~\cite{Gaud06} reactions  described here, were used
to  determine  the  evolution  of intruder  configurations ($N=27$, Sect.\ref{44Ardp})  and locate single-particle
states ($N=29$, Sect.\ref{46Ardp}).

\subsubsection{Study of the neutron single particle states
in $^{47}$Ar \label{46Ardp}\\}

\begin{figure}
\begin{center}
\includegraphics[width=8cm]{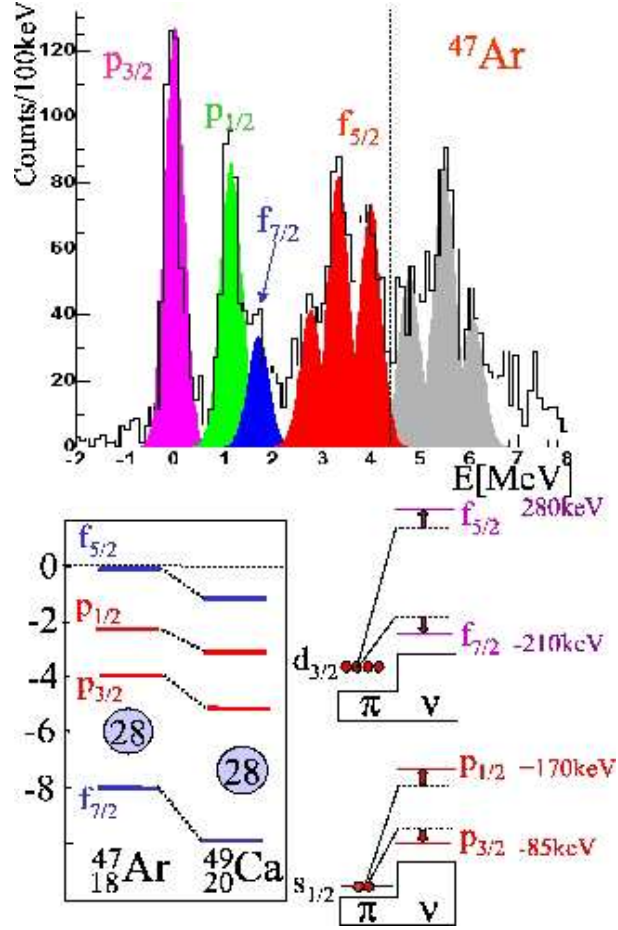}
\caption{(colour online)  Top:   Excitation  energy   spectrum   of  $^{47}$Ar
  obtained  in  the $^{46}$Ar(d,p)  reaction. The dotted vertical line
indicates  S$_n$.  Bottom left  :
  Neutron  Effective  Single Particle  Energies  (ESPE)
derived  from experimental  data  for the  $fp$
  orbitals  in   $^{47}_{18}$Ar$_{29}$ and  $^{49}_{20}$Ca$_{29}$.
Bottom Right :
  Schematic  view  of  the  proton-neutron  interactions  involved  in
  changing  the $f$  (top) and  $p$  (bottom) SO  splittings. For  the
  $p_{1/2}$  and $p_{3/2}$  orbitals, the  spin-dependent parts  of the
  total monopole interactions  were deduced to be $\tilde{V}_{s_{1/2}
    p_{1/2}}^{pn}$     =      +170~keV     and     $\tilde{V}_{s_{1/2}
    p_{3/2}}^{pn}$ = -85~keV}.
\label{46Ar_results}
\end{center}
\end{figure}

The $^{46}$Ar(d,p)$^{47}$Ar reaction was studied using a pure beam
of $^{46}$Ar  (2$\cdot$10$^{4}$~pps) at 10.2~MeV/u and a
0.38~mg/cm$^{2}$  thick CD$_2$  target \cite{Gaud06}.  The  energy
and angle of the protons were measured using 8 modules of the MUST
detector array~\cite{MUST},  which covered  polar  angles ranging
from 110$^\circ$  to 170$^\circ$  with respect  to  the beam
direction. The  transfer-like products  $^{47}$Ar ($^{46}$Ar)  were
selected  and identified  by the SPEG~\cite{SPEG}  spectrometer in
the case  of a  neutron  pick-up to bound (unbound) states  in
$^{47}$Ar. The measured excitation energy spectrum is  shown in
Fig.~\ref{46Ar_results}. The measurement of the $Q$-value  for the
transfer  reaction  to  the   ground  state of $^{47}$Ar  allowed  a
determination of the  N=28 gap in $^{46}$Ar of 4.47(9)~MeV, which is
330(90)~keV smaller than that in $^{48}$Ca. This reduction occurs as
a result of the removal of  two protons from the $d_{3/2}$ and
$s_{1/2}$ orbits which are quasi-degenerate in energy at $N=28$.

Assignments of the angular momenta $\ell$ and SF (or values of the
vacancy expressed as  $(2J+1) \times SF$) for  the various states
identified in the top  panel of  Fig.~\ref{46Ar_results} were
obtained from a  comparison of the experimental  proton angular
distributions  with DWBA   calculations. From the measured values of
the vacancy  of  the ground state 3/2$^-$ (2.44  instead of 4 for a
closed shell) and  first excited  state 7/2$^-$  (1.36 instead  of 0
for  a closed shell),  it was  inferred  that neutron particle-hole
excitation  are already occurring across $N=28$ in the $^{46}$Ar
ground state.

Measured energies and  SF in $^{47}$Ar (and  those in $^{45}$Ar
~\cite{45Ar})  were  used to  improve  the  accuracy  of the  relevant
monopole  terms as  compared to  the  interaction suggested by  Nummela {\it  et
  al.}~\cite{Numm01}.  With  this new shell-model (denoted SDPF-NR) interaction
the  ESPE  of the  $\nu  f_{7/2}$, $\nu
p_{3/2}$ and  $\nu p_{1/2}$ and  $\nu f_{5/2}$ orbits were  deduced in
$^{47}$Ar    and    compared     with    those    in    $^{49}_{20}$Ca
\cite{Abeg78,Uozu94},    as   shown   in    the   bottom    panel   of
Fig.~\ref{46Ar_results}.   As compared to $^{47}$Ar its isotone $^{49}$Ca
has two additional protons in the $d_{3/2}$ or  $s_{1/2}$ orbitals~\cite{Cott98}.
 The  orbits in which the angular momentum is  aligned ($\ell _\uparrow$)
with the  intrinsic spin, such
as  $f_{7/2}$ and  $p_{3/2}$, become  relatively more  bound  than the
$f_{5/2}$  and  $p_{1/2}$  orbits   where  the  angular  momentum  and
intrinsic  spin are  anti-aligned  ($\ell_\downarrow$).  This  feature
suggests  the existence  of  spin-dependent terms  $\tilde{V}$ in  the
proton-neutron  monopole matrix  elements.  Quantitatively,  values of
$\tilde{V}_{d_{3/2}f_{\downarrow,\uparrow}}^{pn}$ have been determined
to be  $\tilde{V}_{d_{3/2} f_{5/2}}^{pn} \simeq$  +280~keV (repulsive)
and  $\tilde{V}_{d_{3/2} f_{7/2}}^{pn}  \simeq$  -210~keV (attractive)
for the $f_{5/2}$ and $f_{7/2}$  orbitals. In this case, the change of
the SO  splitting for the  $f$ orbitals was tentatively  attributed to
the action of tensor forces  between the $d_{3/2}$ protons and the $f$
neutrons~\cite{Otsu05,Gaud06}. The change of  the SO splitting for the
$p$ orbitals was mainly ascribed  to the removal of a certain fraction
of $s_{1/2}$ protons.  As the  protons in the $s_{1/2}$ orbital occupy
the central part of the nucleus,  the change of the neutron $p$ SO
interaction can arise from  a  depletion in central  density of the  nucleus. This points    to   the central
density dependence   of   the   SO
interaction~\cite{Gaud06,Todd04}, as opposed to what is normally  associated with a surface  effect. A reduction  in the  $N=28$ gap
and  the SO
splitting are further expected  in $^{43}_{14}$Si where there are
six less protons than in  $^{49}_{20}$Ca.
This  global shrinking  of  ESPE will
reinforce particle-hole  quadrupole excitations across  $N=28$ (occuring to some extent
in $^{47}_{18}$Ar).   The  observed shape  coexistence
between   a   spherical  and   deformed   state  in   $^{43,44}_{16}$S
\cite{Gaud09,Grev05}   and   the  very   low   2$^+$  energy   in
$^{42}_{14}$Si$_{28}$~\cite{Bast07} confirm this qualitative statement
about increased quadrupole collectivity  below $Z=18$ derived from the
extrapolation of the ESPE.

\subsubsection{Study of intruder states in $^{45}$Ar
\label{44Ardp}\\}

\begin{figure}
\begin{center}
\includegraphics[scale=0.8]{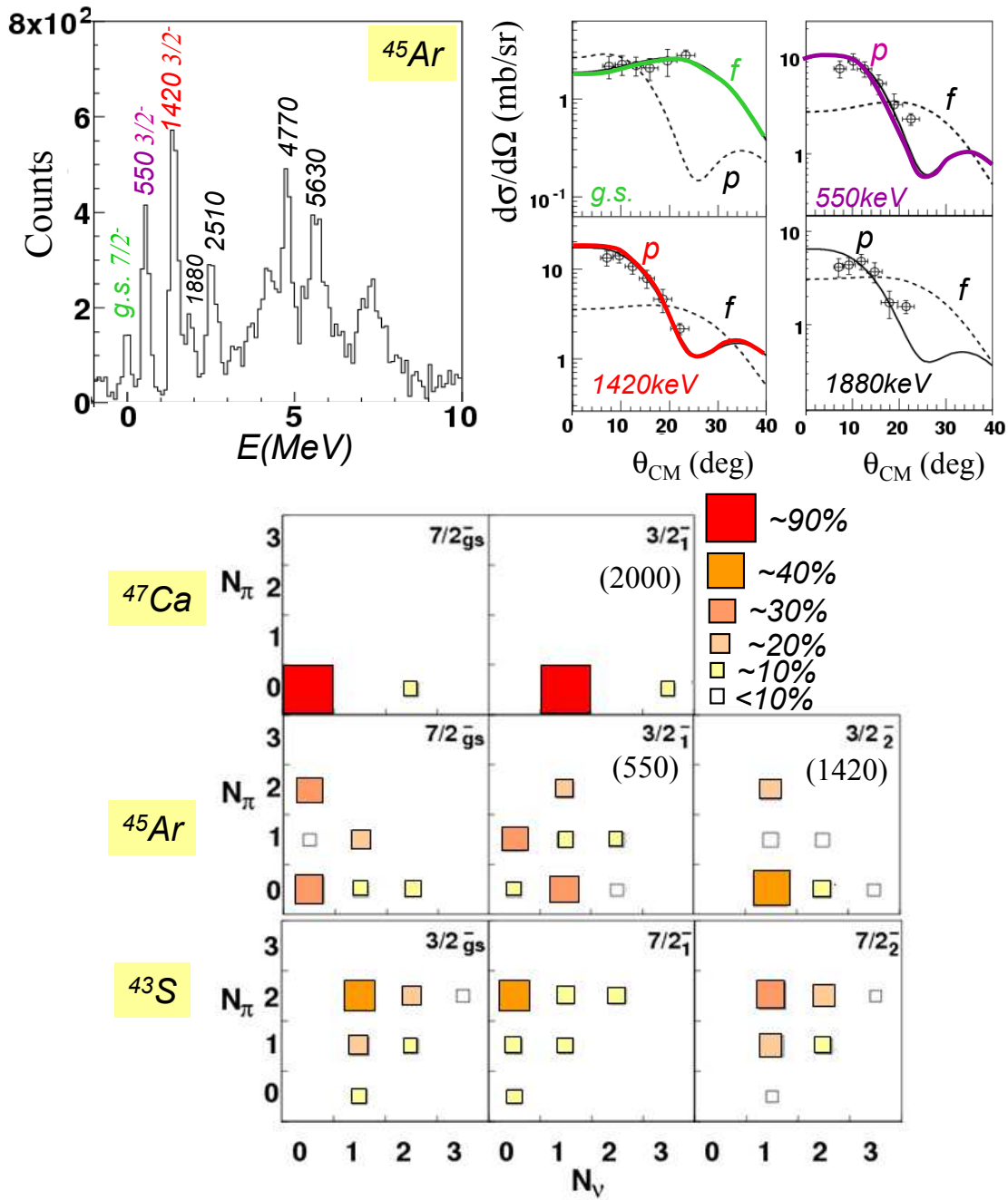}
\caption{(colour online)  Top  Left: Excitation  energy  spectrum of  $^{45}$Ar
  obtained  through   (d,p)  reaction~\cite{45Ar}. Top  Right:
  Angular distributions  corresponding to  the four first  states are
  compared   to  DWBA   calculations.  Angular   momentum   values  of
  $\ell=3,1,1,1$   are   derived    for   the   four   first   states,
  respectively. Bottom  : Configurations of  the ground state
  and  excited states of  the $^{47}$Ca,  $^{45}$Ar and  $^{43}$S are
  compared   in   term  of   neutron   $N_\nu$   and  proton   $N_\pi$
  excitations.}
\label{44Ar_results}
\end{center}
\end{figure}

The $^{44}$Ar(d,p)$^{45}$Ar transfer  reaction study~\cite{45Ar} was made under
similar  conditions to those described above.  Many new states
were populated in neutron transfer and are shown  in  Fig.~\ref{44Ar_results}.
Typical energy resolutions ($\sigma$) of $\sim$120~keV were obtained.
Angular distributions
for the  first four  states in $^{45}$Ar  are also  shown. Unambiguous
angular  momentum  values were  deduced  for  most  of the  observed
levels. In particular,  two 3/2$^-$ states were found  to be populated,
the lower one corresponding to the known 550~keV short-lived isomer Ref.~\cite{Domb03}.

The  evolution  of  the  nuclear  structure  in  the  $N=27$
isotones illustrates how  collectivity develops below $Z=20$.  In
$^{47}$Ca the proton  $sd$  shells  are  completely  filled,  giving
rise  to  wave functions for  the first two states  that have no
component of proton excitation (within  the proton $sd$  valence
space). In  addition, the relatively large $N=28$ gap prevents
neutron cross-shell excitations in the  ground  state.  Therefore,
the  ground  state  wave  function  of $^{47}$Ca is  mainly a
neutron $1h$ configuration  with respect  to the $^{48}$Ca core.
Combining proton  and neutron information, it is found that more
than 90\% of the ground state wave function of $^{47}$Ca contains no
proton and neutron excitations, i.e.  $N_\pi$=$N_\nu$=0
(Fig.~\ref{44Ar_results}). For similar reasons,  the  configuration
of  the  first  excited  state  is  also relatively pure
(Fig.~\ref{44Ar_results}). It  consists of promoting a neutron  into
the $p_{3/2}$ orbital ($N_\nu$=1),  leaving two holes  in the
$f_{7/2}$  orbit, giving rise to a $1p2h$ configuration.

In the   $^{45}$Ar isotone,  the wave functions  are more
mixed (Fig.~\ref{44Ar_results}). This occurs for  two reasons. First, the proton $sd$ shell is
no longer  completely filled at  $Z=18$, as such  excitations can
occur within the $sd$ shells,  leading to a proton configuration which
extends  to  one or  two  proton  excitations ($N_\pi$=1,2).   Second,
neutron excitations  across the $N=28$ gap  are favoured as  the gap is
weakened.  The
ground state  still has a dominant, closed  neutron core configuration,
having $N_\nu$=0. The second  3/2$^-$ state in $^{45}$Ar is  similar
to  the  first 3/2$^-$  in  $^{47}$Ca.  The  first excited  state  (at
550~keV)  contains  a significant  fraction  of  a  recoupling of  the
7/2$^-$  hole with  the  2$^+$  excitation of  the  protons. This  has
implications  in  understanding  the  structure of $^{43}$S.  In
$^{43}$S  an inversion  occurs between  the normal  $1h$ configuration
(which is now the first  excited state) and the intruder configuration
(the ground  state) in which  one or two  neutrons are excited  to the
$p_{3/2}$  orbit. Having  strongly mixed (Fig.~\ref{44Ar_results}) but  similar configurations,
the 3/2$^-$ intruder  state   and  the  second  7/2$^-_2$ state  are
members  of a  rotational band.   The 7/2$^-_1$  has a  more spherical
configuration. This  feature leads to shape coexistence in $^{43}$S
between  the deformed  ground  state and  the  spherical  first
excited state, as discussed in Ref. \cite{Gaud09}.

\subsection{Spectroscopy of light neutron-rich nuclei at the drip line and beyond}

Continued experimental effort over  the last  two decades into
studying neutron-rich nuclei in the lowest part  of  the  nuclear  chart have
revealed   the existence  of  exotic  structures,
such  as halo  states and  new clustering phenomena. On the theoretical  side, ``ab-initio" calculations
are today able to  successfully reproduce  properties of light nuclei
up  to mass  12, starting  from  the fundamental  constituents of  the
nucleus. However,  the treatment  of the continuum  states and  of the
cluster   phenomenon  in  such   approaches  is  not  yet  fully
satisfactory, necessitating  additional and more detailed  data on the
lightest nuclei. The  study of light neutron-rich  nuclei at the limit
of the drip line has benefited from the unique $^6$He and $^8$He beams
produced  by SPIRAL  at  energies  between 3  and  25 MeV/u  (16
MeV/u for $^8$He).  In  particular, the isotopic chain of Helium
was extensively studied by  means of inelastic and transfer reactions,
such  as $^8$He(p,p')$^8$He*, $^8$He(p,d)$^7$He,  $^ 8$He(p,t)$^6$He* and
$^8$He(d,p)$^9$He.  The  most extreme region  of the nuclear  chart in
terms  of  neutron-proton  imbalance  was explored,  with  experiments
dedicated to the search for $^7$H and 4-neutron clusters.

\begin{figure}
\begin{center}
\includegraphics[height=7cm] {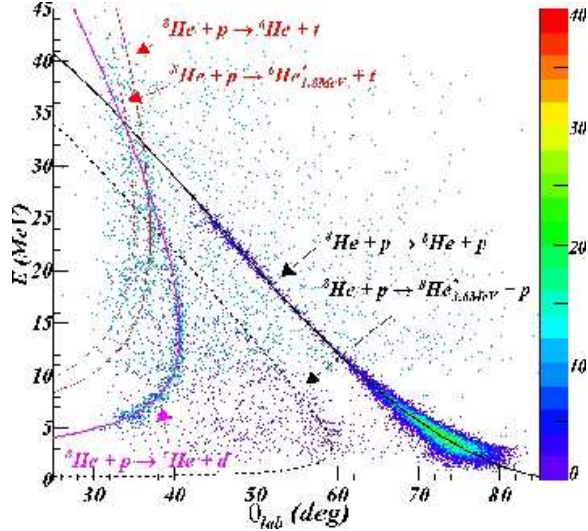}
\caption{\label{MUST2spectr} (colour online)  Energy-angle correlations  for the light  target-like
  recoil, corresponding to elastic  and inelastic scattering of $^8$He
  on a  proton target, $^8$He(p,d)$^7$He  and $^8$He(p,t)$^6$He* where
  the recoil is respectively a proton, deuteron or triton.}
\end{center}
\end{figure}
\subsubsection{Spectroscopy of $^{8}$He \\}

The nucleus $^8$He is the  particle-stable system with the largest  N/Z ratio and
exhibits a neutron skin.  The first experiment performed with a $^8$He
beam from SPIRAL at 15  MeV/u was  dedicated  to the  investigation of  the
properties of the  four-neutron skin of this nucleus,  and to test
nuclear structure  models at the most  extreme values of
isospin. Elastic, inelastic and  transfer cross sections were measured
with the MUST array  and compared to theoretical calculations yielding
detailed  results  on the  structure  of $^8$He  \cite{Ska1,Ska2,Kee}.
Figure~\ref{MUST2spectr} presents  a  typical 2-dimensional  spectrum
showing  the  energy-angle  correlations  of  the  light  partner  for
different reactions  induced by a $^8$He  beam at 15  MeV/u on a
proton  target.   The  kinematical  lines  corresponding  to  elastic,
inelastic  scattering  and  to  one- and  two-neutron  transfer
reactions  can be clearly  identified.   From the  (p,p')  analysis,  the
characteristics of  the excited states of $^8$He  were extracted.
The first 2$^+$ state at  3.62 $\pm$ 0.14 ($\Gamma$=0.3 $\pm$ 0.2 MeV)
and a second  state at 5.4 $\pm$  0.5 ($\Gamma$=0.5 $\pm$ 0.3
MeV) were observed.
Theories that  include the  coupling to  the continuum  are in
agreement with the location of  these two resonant states, while other
microscopic frameworks (no-core shell model or cluster models) predict
higher excitation energies.

Strong coupling  effects of the  (p,d) and (p,t) pick-up  reactions on
the elastic  scattering were  also observed.  Several  coupled channel
calculations were performed using the  Coupled Channel Born Approximation (CCBA)~\cite {Ska1} and a full Coupled  Reaction Channel
(CRC)~\cite {Kee}, using
microscopic or  phenomenological potentials.  In  the CRC calculation,
explicit coupling to the (p,d) and (p,t) channels, including the 2$^+$
excited  state in  $^6$He and  continuum states  in the  deuteron were
taken into account.   A consistent description of the  three data sets
measured  with SPIRAL  beams  together with  data  at higher  incident
energies \cite{Kor} was  obtained. The spectroscopic factors extracted
from    these   analysis    for    the   $^8$He/$^7$He$_{3/2-}$    and
$^8$He/$^6$He$_{0+}$  overlaps  are   in  good  agreement  with  those
obtained  in quasi-free  scattering \cite{Chu}.   The results  of this
analysis  indicate  that   while  the  (1p$_{3/2}$)$^4$  component  is
probably dominant in  the $^8$He ground state, there  is a significant
probability of  finding the  valence neutrons in  other configurations
such as (1p$_{3/2}$)$^2$(1p$_{1/2}$)$^2$,  contrary to the predictions
of the pure jj coupling or of the COSMA model \cite{Kor}.  More recent
data  obtained  using  MUST2  confirm  these  results  and  allow  the
exploration  of  the  spectroscopy   of  $^6$He  at  relatively  high
excitation energy \cite{Mou1}.

\subsubsection{Shell inversion in $^{9}$He\\}

\begin{figure}
\begin{center}
\includegraphics[width=\columnwidth] {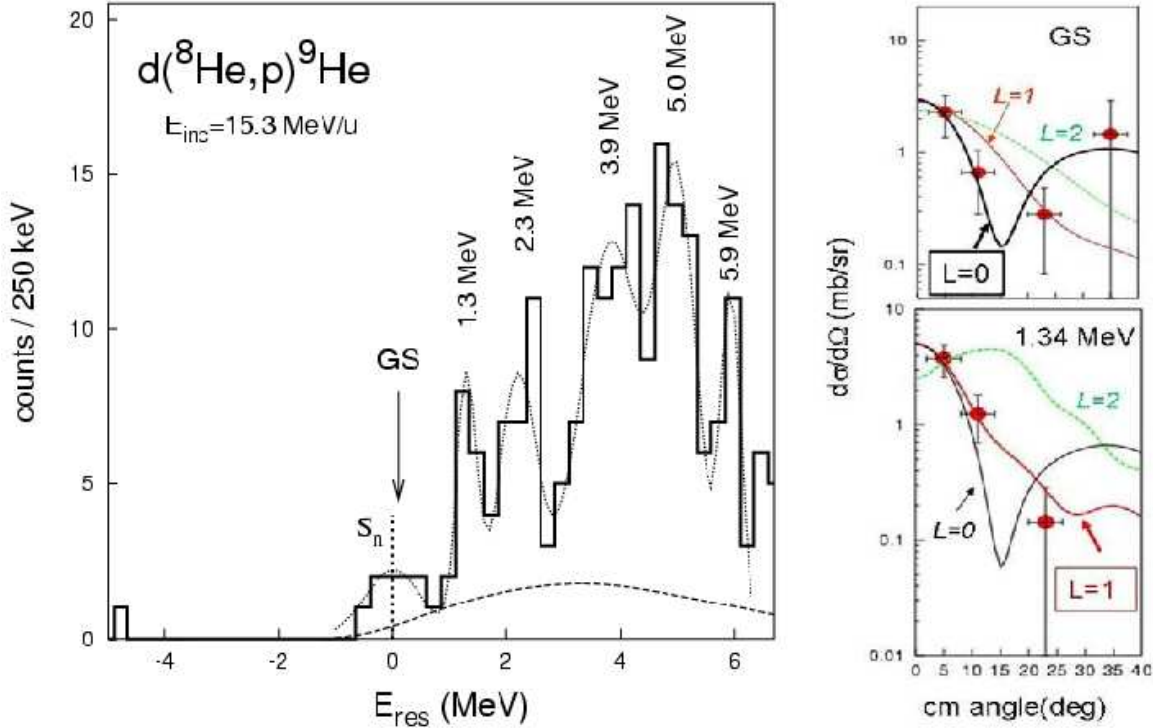}
\caption{(colour online) Left panel:  Excitation   energy  spectrum  of   $^9$He.  E$_{res}$~=~0
  corresponds to the $^8$He+n threshold. Right panel: Comparison of the measured
  angular  distributions  with calculated angular distributions for the states at
  E=0 and  1.34 MeV.
\label{9Hemn}}
\end{center}
\end{figure}

A  dramatic  evolution   of   shell  structure   as  function   of
neutron-proton asymmetry is observed  in the N=7 isotones.  The energy
difference between the s$_{1/2}$ and p$_{1/2}$ orbits is found to decrease going from $^{15}$O
to $^{11}$B, and the level ordering is inversed  in $^{11}$Be, with the
ground  state  being  1/2$^{+}$  instead of  the  expected  1/2$^-$
assignment.  The results  obtained from various studies of  $^{10}$Li find such a  parity
inversion.   Such  an  inversion  can be  understood  considering  the
properties  of   the  spin-isospin  dependent  part   of  the  nuclear
interaction~\cite{Ots}.  In   order  to   search   for  a   possible
p$_{1/2}$-s$_{1/2}$ shell  inversion in  the next N=7  nucleus,
$^9$He, the $^8$He(d,p)$^9$He transfer reaction was measured at a beam energy of  15.3 MeV/u
using the  MUST   array~\cite{Fortier}.  The nucleus $^9$He    was
 previously    studied    using    ($\pi^+$,$\pi^-$),
($^{13}$C,$^{13}$O)  and  ($^{14}$C,$^{14}$O)  double charge  exchange
reactions on $^9$Be~\cite{Seth, Boh}. A narrow resonant state
1.3 MeV above the neutron threshold  was identified as the ground state.
On  the other hand,  results obtained from  studies using the fragmentation of $^{11}$Be~\cite{Chen,Alfalon07}
 were found  to  be  consistent with  the  existence of  a
s$_{1/2}$ ground state  $\sim$0.1 MeV above threshold. The most recent results,
using a knockout reaction, have been obtained at GSI  using a
280 MeV/u  $^{11}$Li beam~\cite{Johan2010}. A transfer $^8$He(d,p)$^9$He reaction studies
at  Dubna \cite{Gol}, showed the presence of two broad
resonances located around  2.0 and 4.5 MeV.  The  analysis showed that
these  could arise from the interference of the 1/2$^-$ resonance with a
virtual 1/2$^+$ state situated close to threshold and also with a 5/2$^+$
resonance above 4.2 MeV.

 The  missing mass spectrum of $^9$He obtained
from  the  $^8$He(d,p)$^9$He reaction study at  SPIRAL is shown in  the  left  panel  of
Fig.~\ref{9Hemn}. This was derived from the energy-angle correlation of the
protons  at backward  angles detected  with the  MUST
array~\cite{Fortier}. Despite  the rather limited statistics, several peaks can
be  identified at  1.34,  2.38,  4.3 and  5.8  MeV. The measured proton  angular
distributions are compared with calculations for neutron transfer
to unbound single-particle states made using the DWBA formalism (right panel of Fig.~\ref{9Hemn})
to determine the corresponding transfered angular momentum and spectroscopic
factor.  The   small  natural  widths  (much smaller than  single  particle
estimates) of $^9$He  states previously observed in charge-exchange
reactions  are  unexpected and  are in  contrast  to  the broad resonances observed in
neighbouring unbound  nuclei.  The  data taken at SPIRAL confirm the existence
of the two narrow low-energy  resonant states (E=1.34 and 2.38 MeV) reported in
Refs~\cite{Seth,Boh},  but  are inconsistent with  the  broad resonances observed  in \cite{Gol}.
Assuming an $l=1$ transition, the measured spectroscopic factor for the
 1.34  MeV resonance  is  much  smaller than  1,  indicating a  strong
 fragmentation  of the  p$_{1/2}$ single  particle strength.  Regarding
the possible s$_{1/2}$  state, the  data suggests
that the $^9$He  ground state is located just  above the neutron threshold
(E=0.1 MeV) with   J$^\pi$=1/2$^+$, in agreement with  the results of~\cite{Chen}.
To improve upon the uncertainties in the angular distributions and $l$ assignment an experiment with improved
statistics using  the MUST2
detector  array has been  performed and is presently under  analysis. The  larger statistics obtained
should help in a better understanding of  the spectroscopy of $^9$He.

\begin{figure}
\begin{center}
\includegraphics[height=6cm] {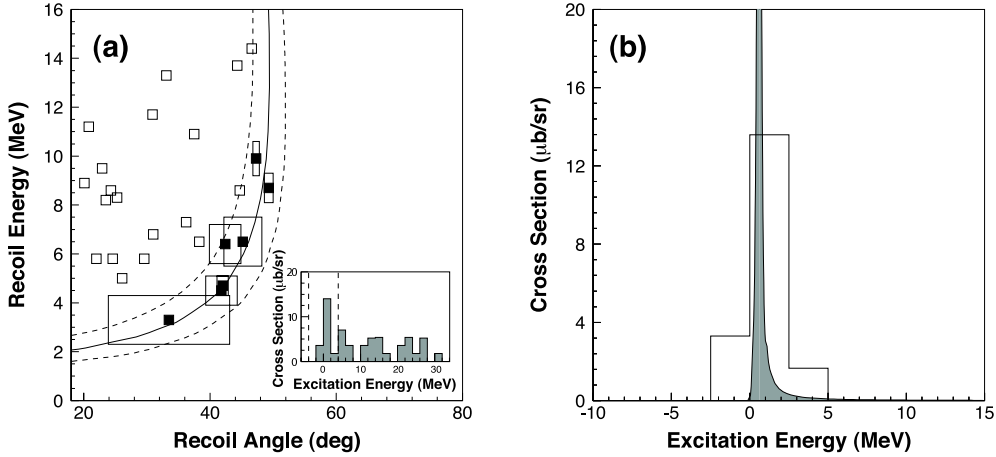}
\caption{a) Range-angle correlations for  the events having a triton
  in the  forward telescopes  and an isotope of  Nitrogen detected
  in the  MAYA gas  volume. The kinematical  line for  the 2-body
  reactions leading to $^{7}$H is indicated by the full line and the
dashed lines correspond to limits of experimental resolution and the width
of the resonance. The latter are also shown in the inset which represents the
excitation energy spectrum for the various events.
b) Excitation energy
  distribution of  the events  identified as $^7$H.   The mass  of the
  $^3$H+4n system corresponds to the zero energy, and the distribution
  is  normalized  to  the  production  cross  section.   The  data  are
  represented by the histogram with a 2.5 MeV binning corresponding to
  the  average  estimated  uncertainty.   The solid  function  is  the
  Breit-Wigner distribution resulting from the fit to the experimental
  events.
\label{7H}}
\end{center}
\end{figure}

\subsubsection{Superheavy hydrogen $^7$H \\}

The system $^7$H represents  the most exotic state  of matter, next
to  that of a neutron star and points towards  the question of  the
number of neutrons that can be held by a single proton. The  first
tentative evidence for the existence of   $^7$H  as   a  resonance
was  obtained   at  RIKEN   using  the $^8$He(p,pp)$^7$H reaction at
61.A  MeV~\cite{Kor}.  A  structure just above  the  t+4n threshold,
superimposed  on a  large background was observed. However,  owing
to the  poor energy resolution (1.9  MeV in the c.m.)  the
properties of the  resonance could not be  extracted. The SPIRAL
$^8$He beams  were used  in two  different experiments to measure
the properties of this resonance using  the missing mass method. The
first experiment used the MAYA active  target filled with isobutane
at 30 mb to  measure the $^ 8$He($^{12}$C,$^{13}$N)$^7$H
reaction~\cite{Caa}.   MAYA was  used to detect the  Nitrogen
recoils  having a total  energy between 3 and 15 MeV. Such  low
energies prevented the  use of a  standard solid Carbon target
where  the  recoils  would  have been  stopped for  a  target
thickness  compatible with  a  reasonable luminosity. The events  of
interest were selected requiring a coincidence between a triton
(decay of  $^7$H) in the forward telescopes  and a Nitrogen recoil
identified in the gas volume of MAYA. The  charge resolution
obtained on the cathode pads  plane  was insufficient to identify
the  mass of  the isotopes   of  Nitrogen produced. Therefore,   the
selected  events could correspond to various isotopes  of Hydrogen.
However, the  kinematics for the different one-proton  transfer
reactions  make it possible to distinguish the various  channels.
Detailed phase  space simulations showed that events corresponding
to the  formation of  $^7$H nuclear  system are located in  a
region of kinematical correlations that can not be accessed by any
other process~\cite {Caa2}. The  existence of the $^7$H ground
state resonance was therefore confirmed with the identification of
seven events where the system was  formed with  a resonance energy
of 0.57$_{-0.21}^{+0.42}$ MeV   above  the $^3$H+4n   threshold  and
a  resonance   width  of 0.09$_{-0.06}^{+0.94}$ MeV
(Fig.~\ref{7H}b).

The second result obtained at SPIRAL  on $^7$H was a by-product of the
search  for  4-neutron  clusters  (described below)  reconstructed  by
selecting the $^3$He  target-like recoils detected with the  MUST array,
corresponding  to the proton  pick-up reaction  $^8$He(d,$^3$He).  The
resulting  spectrum  is  shown  in  Fig.~\ref{4n},  after  subtracting
the contribution  arising from  reactions  with the  Carbon  atoms of  the
CD$_{2}$  target.   The  dotted  line  shows  the  6-body  phase-space
convoluted  with the detection  efficiency. The  existence of  a low-lying
resonance in the  $^7$H system above the t+4n  threshold was confirmed,
with  a   resonance  energy  of   1.56$\pm$  0.27  MeV  and  a  width
$\Gamma$=1.74$\pm$0.72  MeV.  The  discrepancy  between the  resonance
parameters deduced from  the two SPIRAL experiments may  be partly due
to the poor statistics  in these pioneering studies. In addition, one
cannot  exclude that  different states  could  be populated  in  the
(d,$^3$He) and ($^{12}$C,$^{13}$N) reactions.

\subsubsection{Search for neutron clusters \\}

The  search  for  neutron  clusters   has  a  long  history  at  GANIL
\cite{Det}.  More  recently it was triggered  by intriguing
results obtained  in an experiment  studying the breakup  of $^{14}$Be
\cite{Mar}.  In this experiment,  the neutrons were detected using the
DEMON array,  in coincidence with  charged fragments measured  using a
Si-CsI telescope placed around zero degree.
The analysis revealed the presence of some 6 anomalous events in the neutron detectors,
where the deposited energy exceeded significantly the energy deduced from the time-of-flight, assuming the detection of a single neutron. Moreover, these events were found to be in
coincidence with $^{10}$Be fragments. None were found in the other channels. A detailed study of
various effects, including pile up, which could generate such events suggested that the signal
was consistent, at the 2-sigma confidence level, with the detection of a bound
tetraneutron~\cite{Mar} or a low-lying resonance in the 4n system \cite{Nigel1}.
 In order to further explore these observations, the same method and detection
systems were used to investigate the breakup at 15 MeV/u of $^8$He (an obvious
candidate for a tetra-neutron search) supplied by SPIRAL. Unfortunately, owing to a series
of technical difficulties in the execution and analysis of the experiment no definite
conclusions could be drawn~\cite{Nigel}.
\begin{figure}
\begin{center}
\includegraphics[height=6cm] {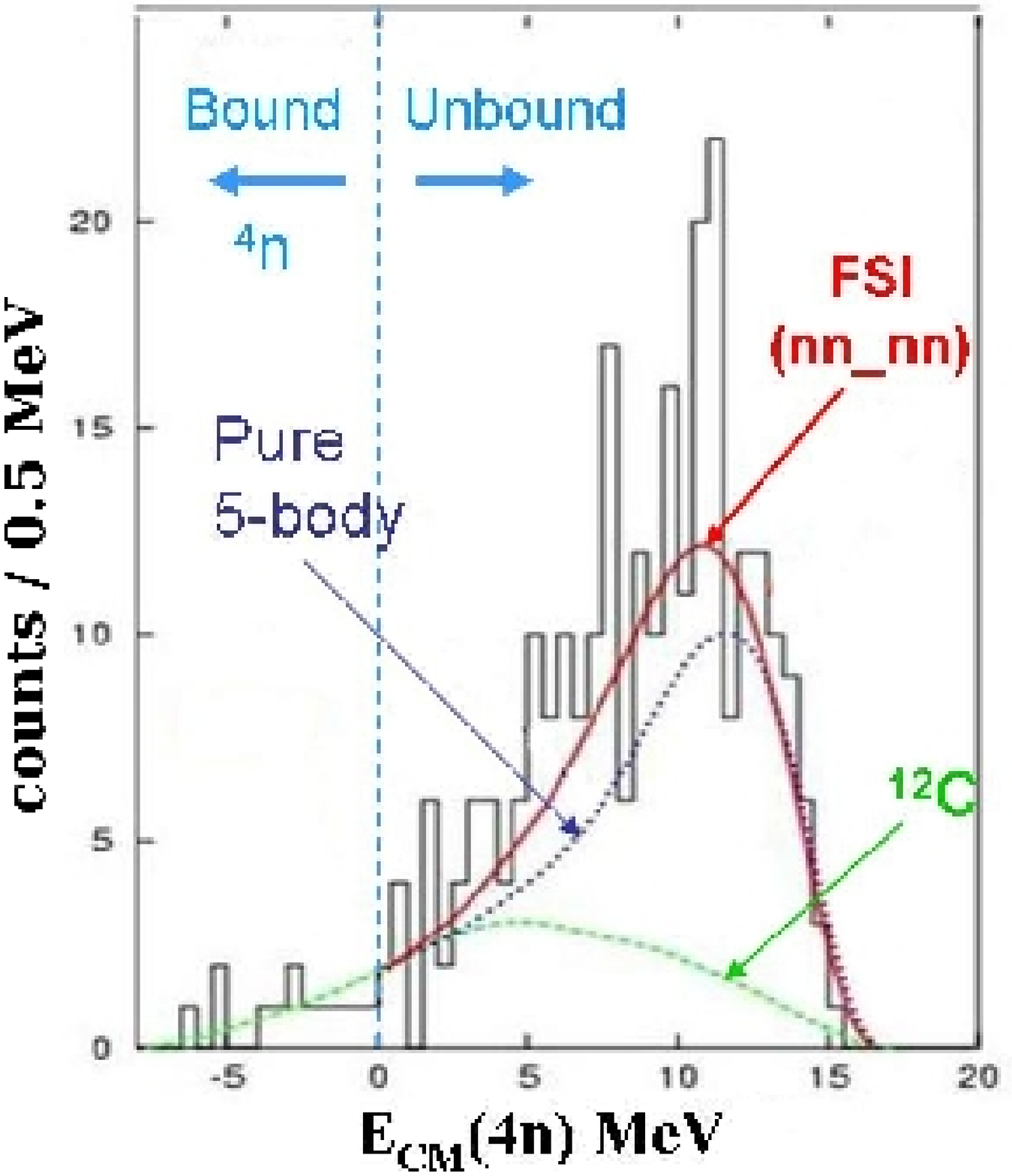}
\includegraphics[height=6cm] {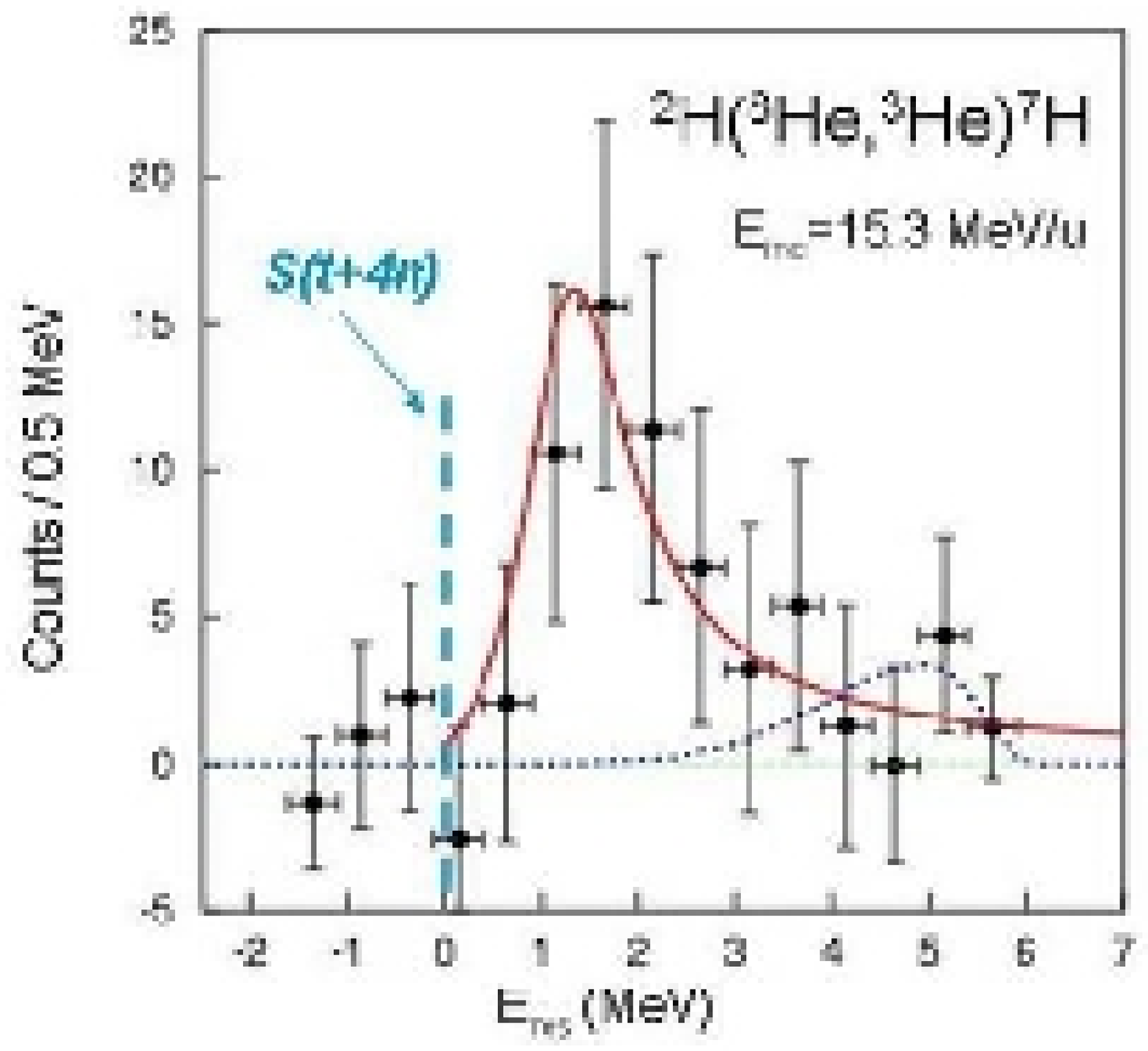}
\caption{(colour online) Left Panel: Excitation energy spectrum  in the 4n system reconstructed
  from the  $^6$Li particles and  gated on a neutron detected in coincidence (see
  text). Right Panel: Excitation-energy spectrum for $^7$H obtained from  $^3$He in the
  same   experiment.   Zero   energy  corresponds   to   the  $^3$H+4n
  threshold. The resonance energy and  width were deduced by fitting a
  Breit-Wigner line shape (solid  line) to the data (see  text). The dotted
  line corresponds to the 6-body phase-space folded with the detection
  efficiency.}
\label{4n}
\end{center}
\end{figure}

In parallel,  the missing  mass method was  exploited to  address this
subject by  studying $^8$He(d,$^6$Li)4n  at 15 MeV/u~\cite{Rich}. The
MUST  array  was  used  to  detect  $^6$Li  produced  by   $\alpha$
transfer. Thin Si layers were added  in front of the MUST detectors
in order to provide an unambiguous identification of $^6$Li particles.
Neutrons  were also  detected at  forward angles  using  thick plastic
detectors  in order  to improve  the reaction  channel  selection. The
excitation-energy  spectrum in the  4n system deduced from  the $^6$Li
energy-angle  correlation   is  presented  in   Fig.~\ref{4n}a.   This
spectrum corresponds  to events  where at least  one neutron  has been
measured  in coincidence  with  $^{6}$Li.  Counts  observed below  the
4n threshold could be  ascribed to the background originating
from $^{12}$C in the CD$_{2}$  target. Thus, no bound tetra-neutron is
observed  within  the  available  statistics.  Above  the  $\alpha$+4n
threshold, a  smooth increase  of the cross  section is  observed. The
data cannot be satisfactorily  reproduced by a pure 5-body phase-space
calculation, which  may indicate the presence  of correlations between
the  4 neutrons  in the  exit channel.   These correlations  have been
investigated  using simple  simulations  of 2n-2n  interaction in  the
final state, and  the result is shown in Fig.~\ref{4n} as the solid
curve.

On the theoretical side, state  of the art $ab-initio$ calculations
lead to the conclusion  that a bound tetra-neutron would  not be
compatible with  our  present  knowledge  of  nuclear  forces
\cite{Pie}.  These calculations however  cannot yet make  reliable
predictions concerning the unbound  case. Solutions of
Faddeev-Yakubovsky equations showed that no  physically observable
four-neutron resonant  states exist~\cite{Laz}.  In conclusion, both
experiment and  theory seem to  agree at  the present time that
bound four-neutron clusters  do not exist.  The situation is less
clear concerning resonances,  such states being more difficult to
observe especially if  they were broad. The results presented in
this section  push the limits  of our  present knowledge of the
nuclear  interaction and structure, and  represent the starting
point for  improving theoretical models, and more generally  our
understanding of nuclear matter.

\section{Reactions and structure studies around the barrier}
The reactions  of heavy ions at energies  around the  barrier are
governed  by a  delicate balance  between the  attractive  nuclear and
repulsive Coulomb  interactions.  These studies have
conclusively demonstrated that the fusion process cannot be understood
as  a simple  barrier  penetration of  structureless  objects, with  a
potential depending  only on the  distance between the centres  of the
colliding nuclei. The interference  between the amplitudes of the many
open and virtual  channels necessitates the need for  treating all the
channels in a coherent manner to understand the reaction mechanisms at
energies around the Coulomb barrier~\cite{Canto}. These amplitudes can
be  tuned   by  changing  the  beam  energy   and/or  using  different
projectile-target combinations.  Radioactive ion beams with their weak
binding, low-lying continuum states and  exotic density distributions
open new avenues to probe and understand the effect of these new and interesting
features on the reaction mechanism~\cite{Nick-R}.  Additionally, reactions  with  RIB can  be  used to  produce
nuclei which  otherwise cannot be  produced in reactions involving stable
beams.  The interest in fusion  studies with radioactive ion beams was
initially  fueled  by   the  conflicting predictions  of  the
influence of weak binding (and thus breakup of the projectile)
on the fusion process.
One approach  treated breakup as causing an attenuation of the flux  in the incident
channel. The transmission coefficients for fusion are  thus multiplied by a breakup survival probability leading
to smaller fusion cross sections. This implies that the two channels, the elastic and the
breakup products, can fuse incoherently leading to complete and breakup fusion, respectively. In contrast to this
intuitive approach, the role of breakup can be considered in a coupled channel formalism like an inelastic excitation
to the continuum. This would mean that the fusing system is a coherent superposition of the elastic and
breakup channels.
It  now understood that  the effects of coupling  to continuum states have to be treated in a coherent manner.
The role of such a coupling is to enhance  (with respect  to a  one dimensional barrier  penetration calculation)
the complete  fusion cross  sections below the  barrier and  reduce it
above the  barrier~\cite{vitturi}. In this section  we focus  on the
exploitation  of  relatively high  intensity   beams  of Borromean systems $^{6,8}$He  at
energies around  the barrier to  understand multidimensional tunnelling
and  to  study  neutron   correlations.   The  exploitation  of  in-beam and off-beam gamma decay of the
residues produced  in reactions  with relatively low intensity RIB  to measure absolute cross sections
is also discussed.

\subsection{Complete reaction studies}
A  comprehensive understanding  of  low-energy  reactions with  weakly
bound  projectiles  requires an unambiguous  identification of the  residues
 produced by different mechanisms and measurements of  the respective  cross sections.
In studies  of fusion  with  weakly bound
neutron-rich nuclei, in addition to the total capture of the projectile
by the target and the breakup of the projectile, capture of  the loosely bound particle(s) by the target
is also important.  Studies performed
at  SPIRAL have shown that the  cross section for the  capture of the
neutron(s)  is  large and  arises  from  a  direct process  (transfer)~\cite{Navin04}. These  events (transfer)  lead to  nuclei
which, in  the case of medium  mass targets, can also  be formed after
evaporation in  a complete fusion  process.  This emphasized  the need
for  identifying  the  mechanism  of  residue  production  (direct  or
compound) in  reactions  involving light  neutron-rich RIB  in order to uniquely
identify the process  of complete fusion.  Similar  issues exist when
the fusion cross  section is obtained from the  measured fission cross
section,  where the  fusion-fission events  have to  be differentiated
from transfer  induced fission \cite{Raabe04}.
This is not  the case with heavier non fissioning compound systems which  decay mainly by the  emission of
neutrons. Thus, in  general, it is
incorrect to equate the total  residue (or fission) cross section with
the fusion cross section.

Results  for  fusion, transfer,  breakup  and  elastic scattering  of
$^{4,6,8}$He on  (medium mass) Cu and  (heavy) Os and  Au targets near
the  Coulomb  barrier  were   used  to  address  the  above  mentioned
issues.  The feasibility  of measuring  small absolute  cross sections
using inclusive  in-beam $\gamma$-ray measurements  with low intensity
ISOL  beams  in conjunction  with  highly  efficient  arrays was  also
demonstrated~\cite{Navin04,Ant09}.  The neutron  transfer process led to the
same residues as those formed in  the decay of a compound nucleus in a
fusion  process.  These  were identified  and separated  using various
measurements of  compound and  direct processes. Neutron transfer
cross sections were measured and found to be larger than those for the
breakup of  the projectile.  This work  was the first to  a) point out
the  importance  of  identifying  and delineating  the  mechanisms  of
residue formation  for understanding fusion  with RIB and  b) demonstrate
the  importance  of  transfer   reactions  with  weakly  bound  nuclei~\cite{Navin04,Nature04,catania}.
With  available  $^8$He   beams  a
complete  study of all  the reaction  channels were  also performed  in the
$^8$He+$^{65}$Cu  systems  at   energies  above  the  barrier.   These
measurements   further  improved  upon   the  sensitivity   of  in-beam
measurements using  re-accelerated ISOL beams studies by  another  two
order of magnitude as compared to the results with $^{6}$He.

\subsection{Neutron correlations in $^6$He}
Neutron-rich nuclei  near the  drip line, especially  Borromean nuclei
(bound  three-body  systems with  unbound  two-body subsystems),  such as
$^{6,8}$He,  $^{11}$Li and  $^{14}$Be, offer  a unique  environment to
study  neutron correlations and  pairing at  low densities. The latter is a
necessary input for nuclear structure models and the study of neutron
stars.   Theoretical  studies indicate showed  that  at  low neutron  density
strong spatial  di-neutron correlations maybe  expected~\cite{hag09}. It
is hoped that the study of the transfer of 1n and 2n  will provide signatures
for the nuclear equivalent of the Josephson tunnelling~\cite{ort01}.  A
related question is whether these weakly bound valence neutrons behave
as independent particles or as  a pair. The ratio of   the 2n/1n  transfer  cross section
could  provide information  about the spatial  correlation of the valence
neutrons. In  $^6$He the  cigar-like configuration, where  the two neutrons  lie on
opposite sides  of the nucleus, should  preferentially populate $^5$He
by  a 1n-transfer    while   the   di-neutron    configuration   should
preferentially lead to  2n-transfer.  The high  intensity of
the  SPIRAL $^6$He  beams  combined  with  the unique  and  high  efficiency
arrays   like  EXOGAM   \cite{Sim00},  the  Neutron  Wall~\cite{Ske99} and
EDEN~\cite{eden}  coupled to  a Si  telescope allowed  the use  of two
independent techniques to explore the spatial configuration of the halo neutrons.

\subsubsection{Transfer reactions in the $^6$He+$^{65}$Cu system \\}

\begin{figure}
\includegraphics[width=\columnwidth]{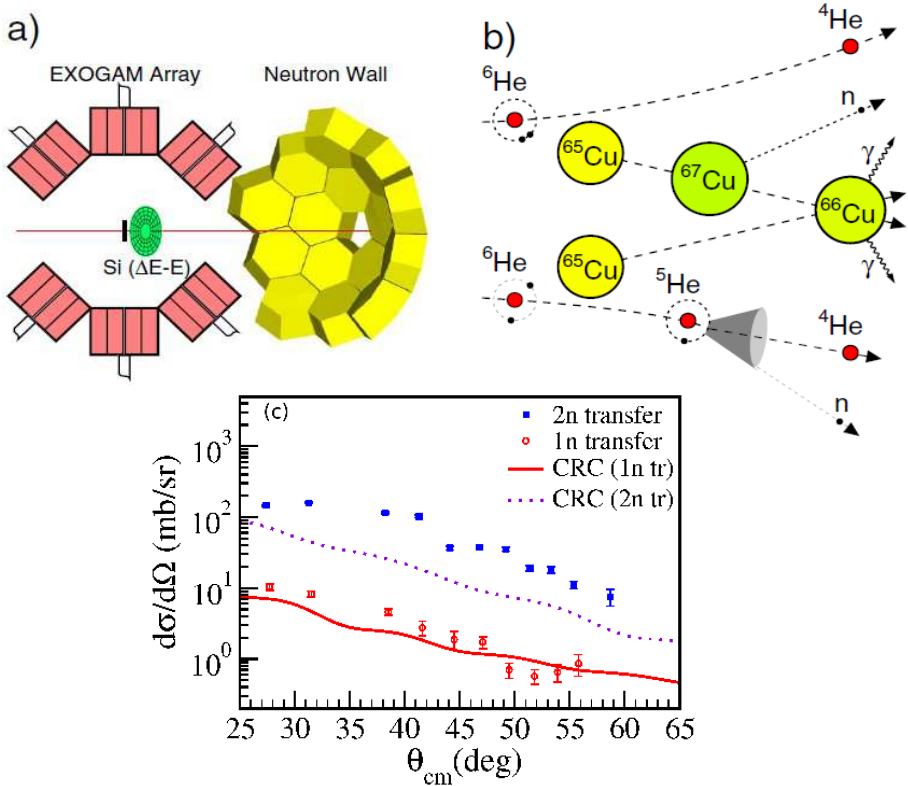}
\caption{(colour online) Schematic  view of (a) the experimental  setup. (b) the reaction
  mechanism for the 2n- and 1n-transfer and  (c) the angular distributions for
  1n- and  2n-transfer. The corresponding CRC calculations  for 1n and
  2n-transfer are also shown.}
\label{6HeCu}
\end{figure}
Triple coincidence measurements,  using intense (4$\times$ 10$^7$ pps)
$^6$He beams, were used to study the neutron correlations by measuring
the 1n  and 2n cross sections  in the $^6$He+$^{65}$Cu system  at 22.6 MeV
\cite{Chat09}.   The   detection  setup  as shown   in  Fig  \ref{6HeCu}a,
consisted of  an annular Si  telescope, the EXOGAM  array \cite{Sim00}
with 11 Compton-suppressed  clovers, and a neutron array  of 45 liquid
scintillator  elements \cite{Ske99}.  The deconvolution of  the $1n$-  and  $2n$- transfer
contributions was obtained from the measured kinematic correlations between the  energies and emission angles
of the $^4$He  particles and neutrons (that  exists for $1n$-transfer
but not for $2n$-transfer) in  coincidence with $\gamma$ rays from the
excited  heavy residue. In the present case, 2n-transfer leads to the formation
of  $^{67}$Cu with an  excitation energy  sufficient to  evaporate a
neutron (isotropically) forming $^{66}$Cu. Owing to the Borromean nature
of $^6$He the final states are  similar in both 1n- and 2n-transfer. In
both cases the  final state consists of a  neutron, alpha particle and
$\gamma$ rays from the excited $^{66}$Cu residue (Fig. \ref{6HeCu}b).  As
the emission  of the  neutron occurs via  to different processes  for 1n
(the neutron arises from the  breakup of $^{5}$He) and 2n transfer
(the neutron  arises from the  evaporation of an excited  $^{67}$Cu) a
kinematic  correlation (between the energies  and emission  angles) exists
between the  alpha particles and  neutrons for 1n-transfer but  not in
the case of a 2n-transfer. Triple  coincidences between neutrons and $\gamma$ rays
from the excited  $^{66}$Cu residues and alpha particles  were used to
deconvolute contributions arising from 1n- and 2n-transfer and
those arising from breakup of the projectile. This work showed the  dominance of the
2n cross sections compared to  1n  transfer cross sections, (Fig.~\ref{6HeCu}c),  thereby
indicating  the  prevalence  of  the di-neutron  structure  in  $^6$He.
Comparison  with  coupled   reactions  calculations  for  the  angular
distributions for transfer,  elastic scattering and fusion illustrated
the important  role played by coupling  to the two  neutron channel in
the reaction mechanism. Lower available
 beam intensities (two orders of magnitude) and the structure
of $^8$He  preclude  such direct measurements with $^8$He beams today~\cite{Lemplb10}.  Therefore alternate methods will be required
to study neutrons correlations in $^8$He using  re-accelerated beams.

\subsubsection{Nuclear breakup in $^6$He+$^{208}$Pb system \\}

\begin{figure}
\includegraphics[width=14cm]{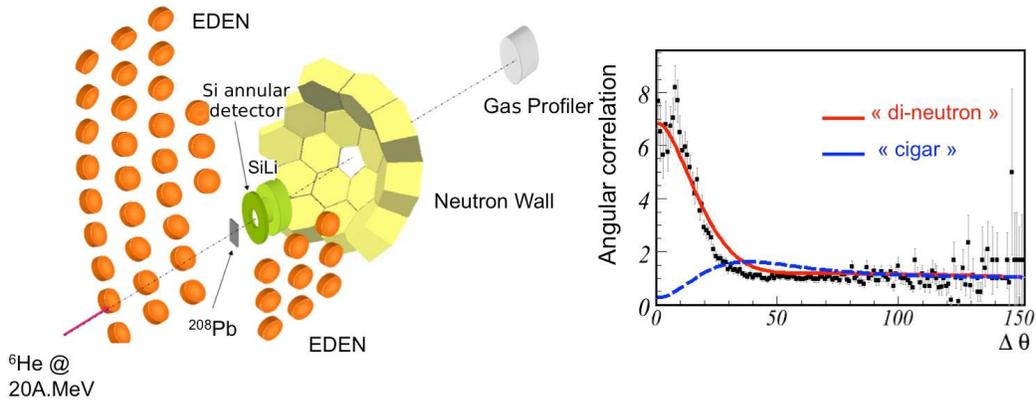}
\caption{(colour online)  Left Panel: Experimental setup for measuring the two neutrons
  produced   in      nuclear   breakup  in   the   $^6$He+$^{208}$Pb
  system.  Neutrons are  detected in  both the  EDEN and  Neutron Wall
  arrays.   Right Panel:  Measured   angular  correlation  between  the  two
  neutrons  emitted in  coincidence  (filled squares).   The data  are
  compared  to microscopic  calculations  where the  two nucleons  are
  assumed to be either in  a di-neutron like configuration (solid red
  line)  using  an attractive  force  between  them  or a  cigar-like
  configuration (dashed blue line) using a repulsive force.  }
\label{6Hepb}
\end{figure}

The  nuclear breakup  mechanism leading  to the  emission of  the two
neutrons  in  the  continuum  was  also used  to  study  the  internal
correlations of the two neutrons in the $^6$He halo.  The measurements
were made using a  15 MeV/u $^6$He  beam on a $^{208}$Pb
target.  By measuring  the two neutrons in coincidence  with the alpha
particle  ejectile  (Fig.~\ref{6Hepb}  left  panel),  the  angular
correlation between the two emitted neutrons initially in the ``halo" was
deduced~\cite{Assie09}.  Theoretical   predictions  based  on  the
microscopic description  of the evolution  of two types  of correlated
systems (cigar  or di-neutron correlation)  suggest a large
sensitivity to  the  initial
correlation  of the neutrons on  the  distribution of the relative
angle  between the neutrons following a nuclear breakup~\cite{AssiePRL09}.
The   right   panel  of Fig.~\ref{6Hepb} compares the experimental observations and theoretical
predictions  showing that  the emission  is compatible  with two
neutrons  in the  ''halo" that  are  spatially close (di-neutron
configuration). Both experiments  discussed in this section, which
employed independent approaches, indicate of a dominant di-neutron configuration in $^6$He.

\subsection{Tunnelling  of exotic  systems}
Historically the alpha particle, the most stable bound helium isotope,
played a  pioneering role  in the first  application of the  theory of
quantum tunnelling.  In this case a pure  barrier penetration by  a structureless
object with  a potential  depending only on  the distance  between the
centres of the interacting systems was considered. The helium isotopic
chain  provides  a  variety  of  properties to  probe  the  effect  of
intrinsic  structure  and  low  binding  energy  on  the  tunnelling
mechanism.  Both  $^6$He  and   $^8$He  have  low  neutron  separation
thresholds and a Borromean  structure.  Contrary to the general trend,
the measured  charge radius of  $^8$He~\cite{Mueller} has been found to be
smaller than that of $^6$He. This was attributed to a more isotropic distribution of
the  four  valence neutrons  around  the  alpha  core. Furthermore,  the
increase of the neutron separation  energy in $^8$He  compared  to $^6$He is  opposite to
the known  behaviour for  any other isotopic chain.

The   helium  isotopic   chain  is  ideal   for understanding the tunnelling       of         composite
objects~\cite{Ber07,Bac06,She08}.   The  ratio  of masses,  $\rho$,  of
interacting  and  non-interacting  components  (with  respect  to  the
Coulomb field) in a composite object,  has been shown to play a significant
role  in   low-energy   reactions,  such as  increasing   the  tunnelling
probability \cite{She08}  (as seen, for  example, in deuteron-deuteron
reactions  at  keV  energies  \cite{Yuk98})  and  creating  cusps  and
resonances  \cite{Ash08}.  The  oscillations  in   the  transmission
coefficient have been shown to  be damped with increasing $\rho$; such
a  result  can  also  be  inferred from  Ref.  \cite{She08}. In  the
neutron-rich helium  isotopes the  valence nucleons outside  the alpha
core can be  treated as the ``non-interacting" analog  given the tight
binding  and point-like behaviour  of the  alpha particle.   The helium
chain offers  a unique progression  of $\rho$ varying from  1 ($^8$He)
and 2  ($^6$He) to the infinite  limit for $^4$He.  In  such a picture
the helium isotopes are expected to behave differently with respect to
their barrier penetration.
\begin{figure}
\begin{center}
\includegraphics[width=1.0\columnwidth]{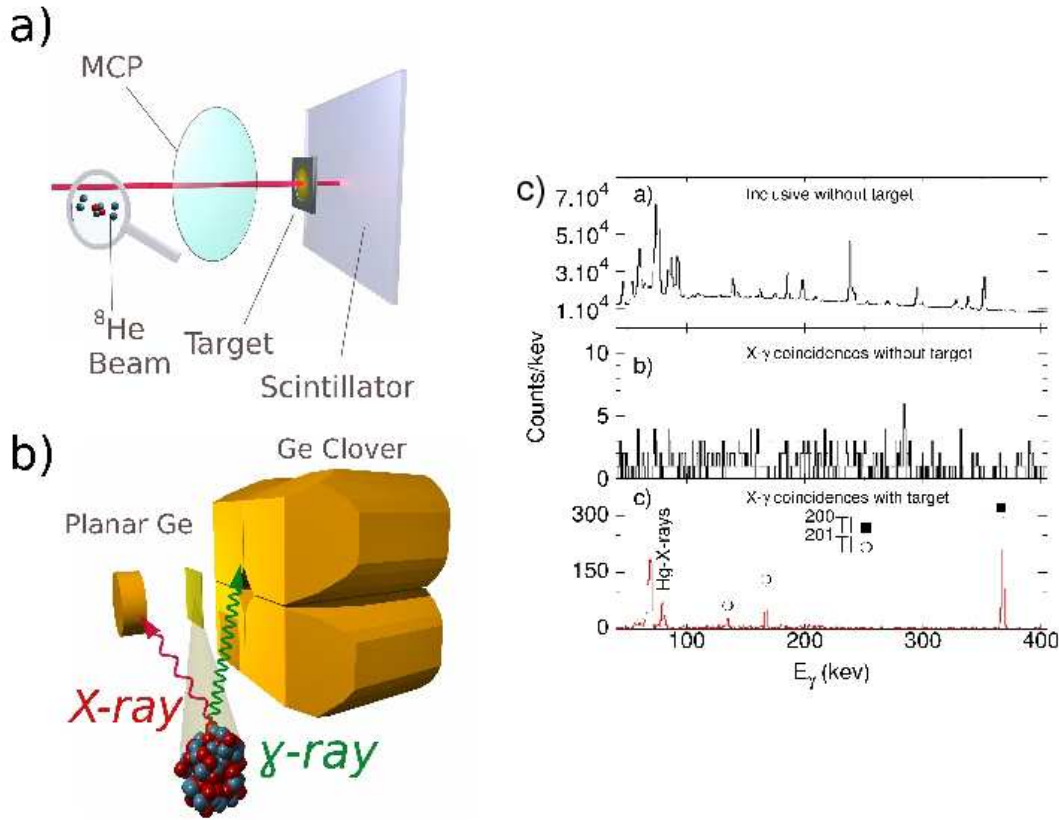}
\caption{ (colour online)  Left: Schematic view of the experimental setup for a) in-beam and  b) off-beam measurements. Right:
  a) $\gamma$-ray spectrum corresponding to a counting time of 210 h,
  without a target and any coincidence condition.  (b) Same as (a) but
  requiring a coincidence condition  on the region corresponding to the  K$_{\alpha}$ X rays of Hg.  (c)
  $\gamma$-ray    spectrum   in   coincidence    with   characteristic
  K$_{\alpha}$ X  rays of  Hg with a  target irradiated at  $E_{lab} =
  22.9 $ MeV.
\label{fig:8He1}}
\end{center}
\end{figure}

The availability of beams  of such nuclei provides  the  opportunity   to  study  the  effect  of  exotic
structures  on the  tunnelling mechanism.   These relatively low  beam intensities  ($\sim$ 10$^5$~pps)
require   employing techniques which
are able  to extract  a weak  signal in the  presence of  a relatively
large  background. The additional  challenge in using in-beam measurements
involving $^8$He beams arise from the large  gamma background  associated  with
the beta  decay of the   beam. Activation techniques,  when applicable,
offer both a  unique identification of the nucleus  (obtained from the
knowledge of the energy  and half-life of the $\gamma$/$\alpha$ decay)
and also  a relatively lower background.  To  overcome the limitations
of a  low signal-to-noise ratio,  a  selective  and sensitive
method  was  designed  to  access  the fusion  cross  sections.   This
involves the  simultaneous measurement of X and  $\gamma$ rays emitted
in electron  capture decay of the  evaporation residues~\cite{nim}.  A
schematic view  of  the  setup  is
provided in  Fig. \ref{fig:8He1}.  The  gain in
sensitivity resulting  from the present method, compared
to an inclusive  measurement, was found to be $\sim  3 \times 10^4$ in
the  region   of  interest.    A  typical  $\gamma$-ray   spectrum  in
coincidence  with   characteristic  K$_{\alpha}$  X  rays   of  Hg  at
$E_{\rm{lab}} =  22.9$~MeV illustrating the selectivity  of the method
is also shown in the figure.  This large gain in sensitivity can be put
in  perspective  by noting  out  that  the  smallest cross  sections
measured in this  work using re-accelerated RIB, when  scaled  by the
million times larger intensities available with stable beams, is  comparable to the
current  measurement limits  in nuclear  physics.
\begin{figure}
\begin{center}
\includegraphics[width=0.5\columnwidth]{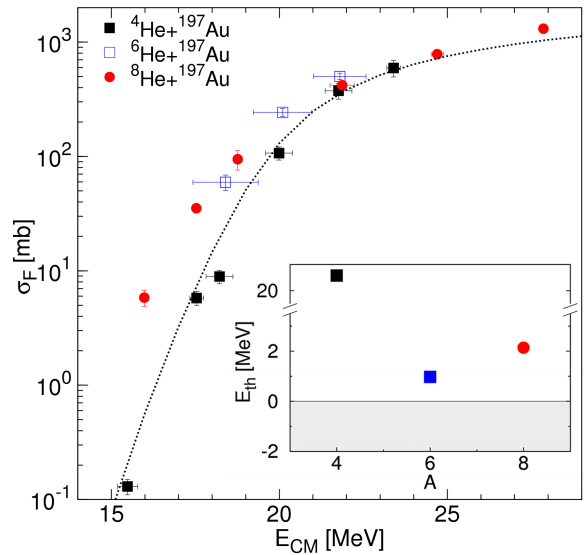}
\caption{(colour online)   Measured fusion cross section as a function of centre-of-mass
  energy    $E_{\rm{CM}}$   for  the   $^{4,6,8}$He~+~$^{197}$Au   systems~\cite{alpha,yuri,lem09}.
  The dotted  line shows a one-dimensional
  barrier penetration calculation  for $^4$He~+~$^{197}$Au.  The inset
  shows  the  lowest threshold  energy  for He isotopes.
\label{fig:8He2}}
\end{center}
\end{figure}

The measured  fusion cross  sections for
Helium  isotopes  with  $^{197}$Au~\cite{lem09,yuri,alpha} are shown in Fig.~\ref{fig:8He2}. The  good  agreement
between the  calculated and measured  fusion cross section  for $^4$He~\cite{alpha}
reinforces its point-like behaviour.  At energies below the barrier the
fusion  cross  sections,  for  $^8$He  and  $^{6}$He~\cite{yuri}  are
surprisingly similar  and, as expected,  are much larger than  for $^4$He~\cite{alpha}.
A  loosely bound system  with a subsystem that  does not
feel  the barrier can more  easily restructure  during  the dynamical
process of fusion,  emphasizing the role of a  flexible intrinsic wave
function  that  can  adiabatically  readjust  in a  slow  process  and
increase the barrier penetration.  The  observed  similarity  of  the  low-energy
results for  $^8$He and $^6$He may  indicate the role  of higher order
processes  with  neutron-pair  transfer  preceding fusion.   The  most
accurate  and complete  measurements of  $^8$He+$^{197}$Au  fusion and
transfer~\cite{lem09} seem  to indicate  that for a  loosely bound  but essentially
isotropic system  like $^8$He, it turns  out to be  easier to transfer
part of the  neutron excess in a peripheral  reaction than to readjust
the outer  skin of the system and  tunnel as a whole,  a process which
occurs  only   in  a  narrow   region  of  small   impact  parameters.
Understanding  the intriguing  behaviour  of the  helium isotopes  may
help  possible future  applications in  the production of  super heavy
elements and the study of decoherence effects in open quantum systems~\cite{hinde10}.
The improved sensitivities demonstrated in this work using radioactive ion
beams  may also permit measurements which were earlier not possible.

\subsection{Nuclear structure studies using $\gamma$-ray spectroscopy \label{coulex}}

Various techniques  such as  Coulomb excitation and  fusion-evaporation are
used to  produce and  characterize excited states  in nuclei  far from
stability using reactions around  the Coulomb barrier.  These measurements were made using EXOGAM
and when necessary in conjunction with auxiliary detectors including VAMOS,
charged particle  detectors  and  neutron  detectors.

\subsubsection{Coulomb excitation to probe nuclear shapes \\}
Coulomb excitation at energies below the Coulomb barrier, is a purely
electromagnetic  process whereby  multiple step  processes can populate relatively
high spin  states.  The  excitation probabilities are
to  first order proportional to the square of the transitional matrix
elements  (B(E2)  values).   The  second  order  reorientation  effect
introduces a  dependence on the  diagonal matrix element, allowing for the determination of
the static  quadrupole moments and  their signs~\cite{Gorgen}.  Thus,
Coulomb excitation  studies are  sensitive to the  distinction between
prolate and  oblate shapes. States of prolate and oblate shape are likely
to coexist within a narrow  range of energy and are expected to be strongly
mixed. This feature of coexistence has been shown to arise from the competition between
large prolate or oblate proton and neutron shell gaps at $N=34,36$ and
$38$      in      the     ($Z=36$,      $N      \simeq     Z$)      Kr
isotopes~\cite{Bend06}.   The first experimental  evidence   of  such   a  shape
coexistence in the  light Kr isotopes was obtained from the observation
of low-lying excited states and their measured electric monopole decay
strength  $\rho  (E0)$. Additionally,  the  mixing  between two  0$^+$
states  induces  a distortion  in  the  regular  sequence of  energies
characterizing rotational  bands of  oblate or prolate  rotors.  This
led   to    the   proposition~\cite{Bouc03} that, the  ground states of $^{76,78}$Kr are prolate,
the prolate and oblate  shapes are strongly mixed in  $^{74}$Kr and
the  ground state configuration  of $^{72}$Kr  is mainly  oblate.  The
experimental program  carried out at SPIRAL aimed at  determining this
evolution in   shapes  in   the  neutron   deficient  Kr
isotopes. The  EXOGAM array  of Ge clover  detectors was  coupled  to a
highly segmented annular Si detector  to detect the scattered
particles in  coincidence with the  $\gamma$-ray transitions following
the de-excitation of states populated in Coulomb excitation.

\begin{figure}
\begin{center}
\includegraphics[width=\columnwidth]{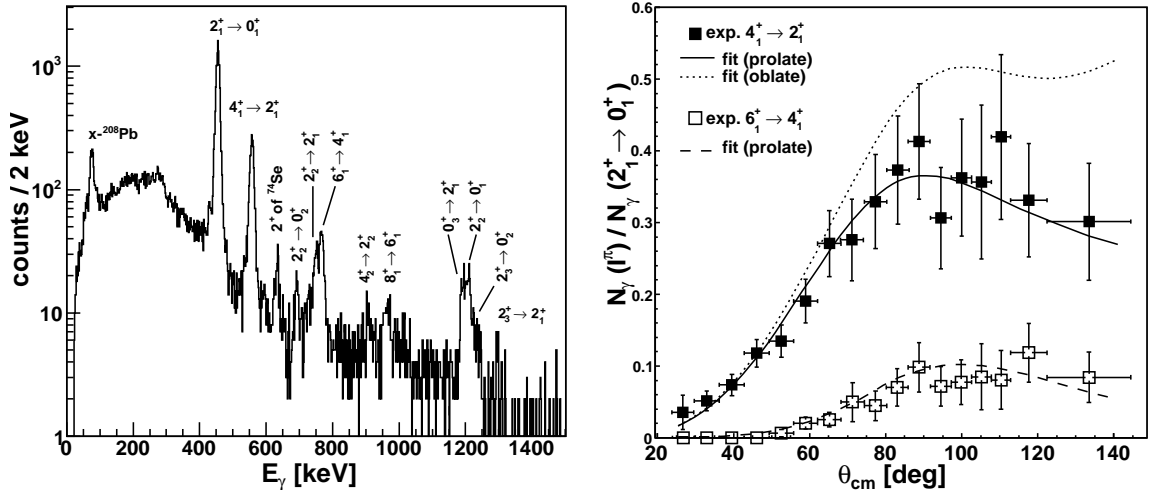}
\caption{Left Panel:   $\gamma$-ray  spectrum  obtained   for  the
  Coulomb  excitation of  $^{74}$Kr.  Right Panel: The  $\gamma$-ray
  yields for the 4$^+  \rightarrow$ 2$^+$ and 6$^+ \rightarrow$ 4$^+$ transitions,
  normalized to the  2$^+ \rightarrow$ 0$^+$, are shown  as a function
  of the centre of mass scattering angle. The full line corresponds to
  a  fit obtained  with  the GOSIA  code~\cite{GOSIA}, for  a prolate ground  state of  $^{74}$Kr.
  To illustrate the sensitivity  of the method for determining the sign of the quadrupole moment, the
is result of the calculation for an  oblate shape (dotted
  line) for the
  ground state.}
\label{Kr_results}
\end{center}
\end{figure}

High quality beams of $\sim$ $5\times$10$^5$ and 10$^4$ pps were used for the
study of $^{76}$Kr and  $^{74}$Kr nuclei, respectively.  The gamma-ray
spectrum   obtained  for  $^{74}$Kr    exhibits  many
transitions and is  shown  in Fig.~\ref{Kr_results} (higher statistics was
obtained  in the case  $^{76}$Kr).  The  excitation  probabilities of  several
transitions were extracted from  the $\gamma$-ray yields as a function
of the scattering angle of the nuclei~\cite{Clem07}. Both transitional
and diagonal matrix elements  were determined by fitting the observed yields.
Negative  values were found for the spectroscopic quadrupole moments  of the first 2$^+$ states of
$^{76}$Kr and  $^{74}$Kr, indicating  a prolate shape.  The quadrupole
moments of the  $2^+_{2,3}$  states were found to be positive and the corresponding shape was interpreted
to be oblate~\cite{Clem07}.  This  result represents  the
first  direct proof for  a prolate-oblate  shape coexistence  in these nuclei. Comparison  with   calculations beyond
the static mean-field approach  emphasized the importance of including
the  triaxial  degree  of freedom  to  describe  the  shape
coexistence  in these  nuclei~\cite{Bend06,Giro09}.  The  present work  was also the first exploitation of
the reorientation effect with   radioactive ion beams. Other  transitional  matrix  elements  between
low-lying states  were also extracted, providing  information about the
mixing of  their wave functions.  The nucleus $^{72}$Kr  is expected to  be one of
the few nuclei  having an oblate ground state, however its study is presently not possible with the available intensity.

Similar studies  undertaken were made in  the case of $^{44}$Ar~\cite{Arcool} in
order to gain an insight into  the weakening of the N=28 shell closure
and the development of deformation  in this region.  The  first and the second 2$^{+}$  states in $^{44}$Ar
were populated in Coulomb excitation  on $^{208}$Pb  and $^{109}$Ag  targets  at beam
energies of  2.68 and 3.68  MeV/u respectively.  B(E2) values between  all observed
states  and the  spectroscopic quadrupole  moment of  the  first 2$^+$
state were  extracted from  the differential Coulomb  excitation cross
sections. These results indicated  a prolate shape for $^{44}$Ar, demonstrating
the onset  of deformation  for a nucleus  already two protons  and two
neutrons     away     from     doubly    magic     $^{48}$Ca.
Calculations were performed  with the  Gogny  D1S  interaction  for $^{44}$Ar  and
neighbouring  nuclei  using   two  different  approaches:  the  angular
momentum  projected  generator  coordinate  method  considering  axial
quadrupole deformations and  a five-dimensional approach including the
triaxial  degree   of  freedom.   The  experimental  values  were compared with new
Hartree-Fock-Bogoliubov  based configuration mixed shell-model calculations and
also with  shell-model  and relativistic mean-field calculations. The calculations suggest a
$\gamma$-vibrational character of the 2$+_2$ state in $^{44}$Ar~\cite{Arcool}.

\subsubsection{Characterization of nuclei away from stability \\}

Fusion  and multi-nucleon transfer reactions using radioactive
ion beams  provide an alternate way to produce nuclei  which are
otherwise  difficult   to  produce  using  stable  beams.
The low beam intensities and radioactive  decay of
the beam make  inclusive  measurements with RIB challenging.
Here we  discuss some of  measurements that have been carried out.

The first experiment with the EXOGAM array,  operated in  a stand alone
mode,  used four  detectors (in a ``cube'' geometry)
to  address the interplay between  vibrational and multi-particle
modes of  excitation in  neutron rich-nuclei  around the  doubly magic
Pb. These  nuclei are difficult to study  at high
spin due  to the  restricted possibilities with  combinations of  stable beams and  targets.  To
overcome  these constraints  studies  were carried  out  using a $^8$He beam
on a  thick $^{208}$Pb  target   to  investigate states  in
$^{212}$Po and  $^{213}$At~\cite{Garn05}.  The measurements performed at a beam
energy of 28 MeV with an intensity of 2$\times$10$^5$ gave access to states up to a relatively  high  spin (14$\hbar$) in   $^{212}$Po.
An  increased
relative population  of states  up to 12$^+$  was observed  when compared
with earlier  work using a  $^{208}$Pb($^9$Be,n)$^{212}$Po reaction at
48 MeV.   Evidence for  a previously  unreported transition  of  69 keV
corresponding to  13$^−$ $\rightarrow$ 12$^+$ was found.  States up to
J$^\pi$ =  (39/2$^−$) were also  observed in $^{213}$At.

Gamma  spectroscopy studies were  also made near the  proton drip  line of  the light  rare earth elements.
In particular, neutron deficient  rare-earth nuclei around mass  130
are of  great interest  since highly  deformed  prolate ground
states are predicted to exist in this region. These nuclei, with neutron
numbers  lying midway  between the  N= 50  and 82  shell  closures and
proton numbers away from the Z = 50 shell closure, are predicted to
show  very large          ground-state        deformations        of
$\beta_2$$\sim$0.4. Experimentally it was possible to populate and study with stable  beams
nuclei around A$\sim$130 only near, but not at the maximum of this predicted deformation.
With  this motivation,  studies with
fusion evaporation  reactions using a $^{76}$Kr beam, with intensities
of  5~$\times$~10$^5$ pps  and an  energy of  4.34 MeV/u  on  a $^{58}$Ni
target were made~\cite{Petri06}.  This  experiment used the
  EXOGAM array in conjunction with VAMOS and  the charged  particle  array
DIAMANT~\cite{daimant} to enhance the selectivity of the relevant nuclei.
The  strong 4p  exit  channel populating  $^{130}$Nd,  was studied  up  to
relatively  high spin and  linear-polarisation  measurements were
also carried  out. These  early experiments  demonstrate the  potential of using fusion
evaporation with   low intensity RIB to populate and characterize states otherwise difficult to reach.

Multi-nucleon  transfer reactions  at energies around the  Coulomb  barrier
allow to  investigate among others  the nucleon-nucleon
correlation in  nuclei, the transition  from the quasi-elastic  to the
deep-inelastic  regime  and channel  coupling  effects in  sub-barrier
fusion reactions~\cite{Nanni}.   These reactions  can also be  used to  produce and
characterize nuclei  far away  from stability.  Compared  to reactions
with stable  beams where  the main transfer  channels are  dominated by
pick-up of  neutrons and stripping of protons,  those with radioactive ion beams
open possibilities to populate nuclei by  both the pick-up and stripping
of  protons  and  neutrons~\cite{dasso94}. Thus  not  only
neutron-neutron   and   proton-proton   correlations  but   also   the
neutron-proton correlations can be studied simultaneously. The relevant information can be obtained from the
observed population pattern of  specific final states.

 Studies of multi nucleon  transfer were made
using beams  of $^{24}$Ne on  a  Pb  target at 7.9 MeV/u~\cite{gevona}.  The
light transfer products were detected  and identified in VAMOS and the
coincident gamma rays were detected  using EXOGAM in a close packed geometry.
Angular  distributions for  the  elastic and  inelastic channels  were
measured, together  with distributions for the  $^{23}$Ne and $^{23}$F
channels.  The relatively low  beam intensity of 10$^5$~pps limited the
population  of more  exotic channels.  Angular distributions of a few channels
were measured and found to be in rather good agreement with predictions from a semiclassical model.
This work points to the potential of using multi-nucleon transfer a  probe of  nuclear structure,
with higher beam intensities.

The  deexcitation of  hot  nuclei is  generally understood  using a  statistical
model. One of the important inputs is the knowledge of the level densities
of the nuclei  before and after particle emission,  in particular the
level density  parameter {\bf\it a}.   To predict the decay  of nuclei
far from stability requires an  understanding of the evolution of the
level density. With this
motivation measurements have also been made using SPIRAL beams to study the
variation of the level density  parameter as a function of isospin for
nuclei  around A=100.   The level  density parameter  was investigated
through the measurement of the evaporation residues using VAMOS in conjunction
with characteristic K X-rays  (for In nuclei)~\cite{Morjean} and
from the measurements of  charged particles with the INDRA detector
array~\cite{INDRA} (Pd   nuclei)~\cite{Nicolas}. The  data  are presently
under analysis.

\section{Nuclear Astrophysics}
Radioactive ion beams play an important role in  understanding how nuclear  processes influence astrophysical
phenomena.  At SPIRAL taking advantage of the available high-quality post-accelerated
beams  two particular  aspects have been  addressed: (i) the role of unbound nuclei
in explosive  combustion of Hydrogen and  (ii) the neutron  capture cross
section  involving an  unstable nucleus.
Direct measurements of cross sections at low energies are also planned in the near future.

In   explosive astrophysical   environments   such   as  X-ray   bursters,   dripline
neutron-deficient  nuclei  can be  strongly  synthesized  by fast  and
successive capture  of protons. This leads to an  equilibrium stage where
the  capture  of  another  proton  is followed  by  the  instantaneous
emission  of  a  proton  by the  neighbouring  proton-unbound nucleus.
Such a  waiting  point  can be  bypassed  by a  two-proton
radiative  capture reaction  if the  final nucleus  is bound.
This is possible     in     several      reactions,     for     example
$^{15}$O$(2p,\gamma)^{17}$Ne   and  $^{18}$Ne$(2p,\gamma)^{20}$Mg.
The  exotic   two-proton  capture  reaction  was   first  proposed  by
G\"{o}rres    \emph{et   al.}~\cite{Goerres:1995}.
These alternative paths can be calculated using the properties of the
intermediate  unbound  nuclei (e.g.  $^{16}$F  and  $^{19}$Na).  Calculated reaction rates for two-proton
capture were  found  to  be significant only for  astrophysical   environments  at   very  high
densities.  The properties  of a large number of low lying  states of  the relevant unbound
nuclei are presently unknown.  Hence an experimental  program  was started  with  the
objective to measure these  properties  with a high  energy resolution and dedicated techniques to perform
resonant  elastic scattering were  developed. These were first applied to  measure the properties of
the low  lying states  in  $^{19}$Na and  $^{16}$F.
As mentioned earlier the  beam energies provided  by SPIRAL are ideal  for transfer
reactions (Sect.\ref{transfer}). The  (d,p) transfer reaction has been
used to  determine the important  observables (Q values, excitation energy
 and  spectroscopic   factors)  required   to   determine  the
(n,$\gamma$) radiative capture rate in the case of $^{46}$Ar.  This nucleus
plays a pivotal role for understanding the reduction of $^{46}$Ca with
respect  to $^{48}$Ca  in  neutron capture  - $\beta$-decay  scenarios
\cite{sorl03,Gaud05}.  The results of these measurements are  presented in the following.

\subsection{States in  $^{19}$Na}

\begin{figure}
\begin{center}
\includegraphics[width=0.5\columnwidth]{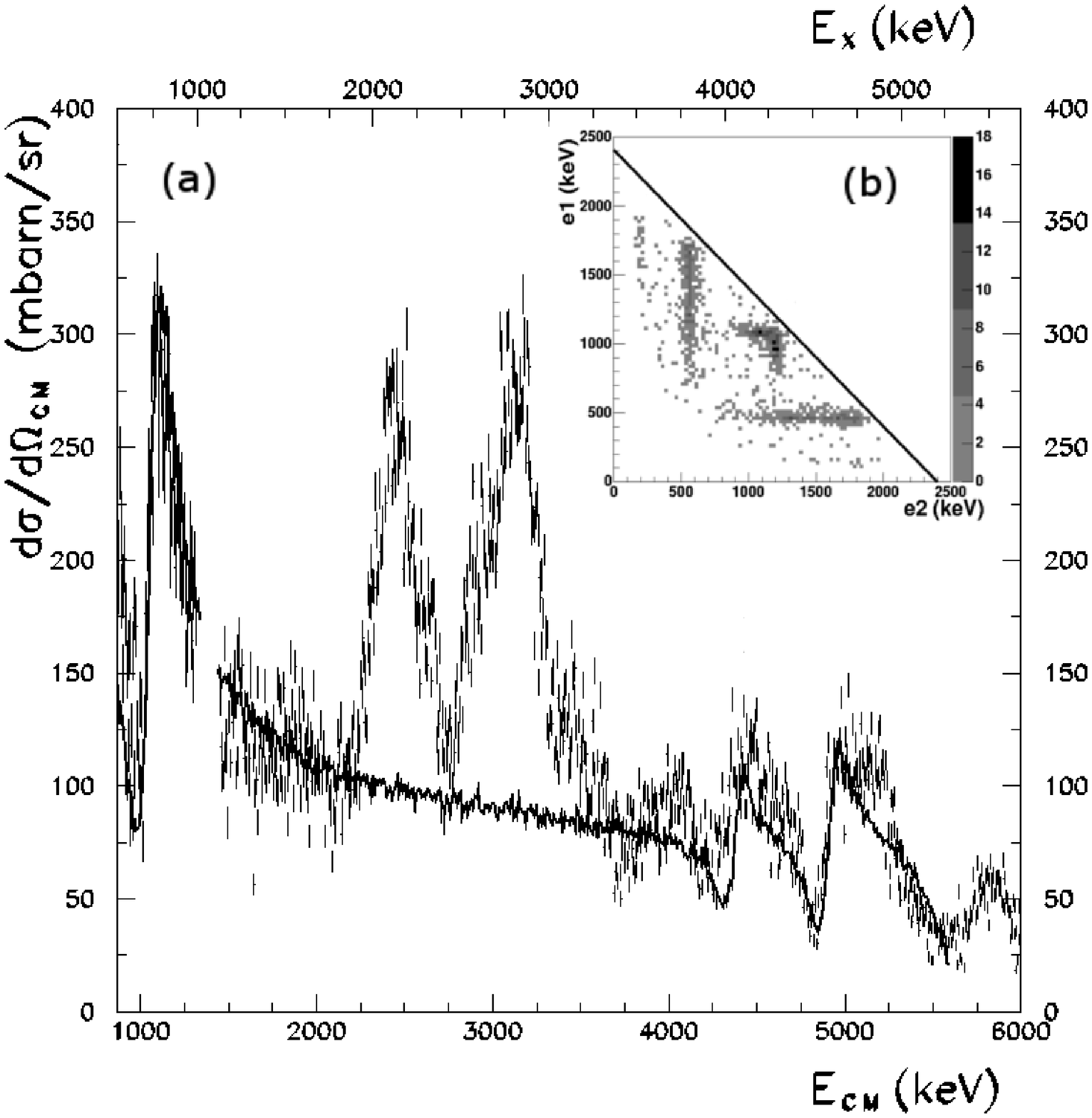}
\hspace{0.5cm}
\includegraphics[width=0.5\columnwidth]{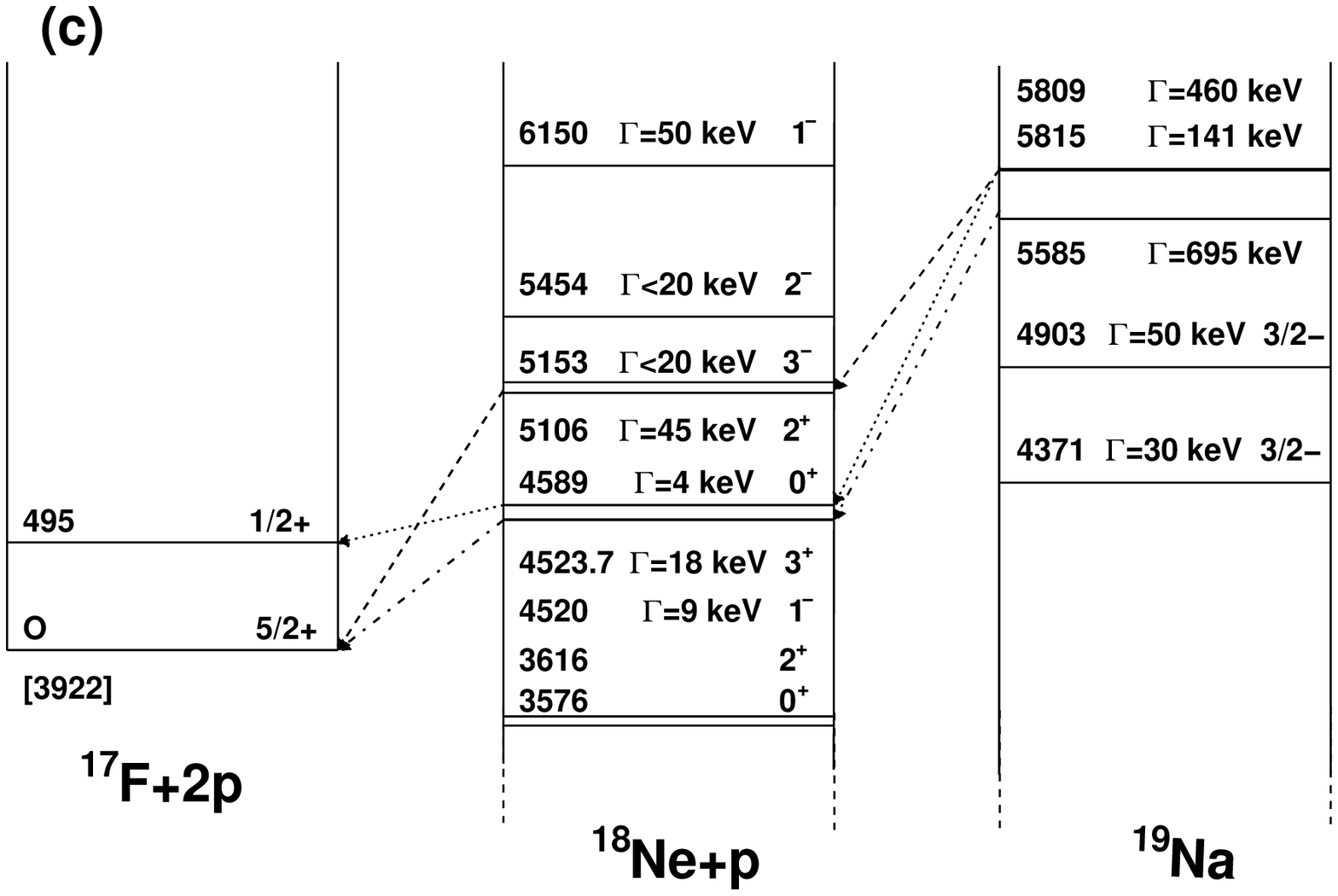}

\caption{\label{fig:na19}(a)     Excitation    function     for    the
  H($^{18}$Ne,p)$^{18}$Ne reaction measured at 180$^{\circ}$ in the centre
  of mass  (C.M.). The  line corresponds to an  R-matrix calculation.
  (b) Two  proton emission H($^{18}$Ne,2p)$^{17}$F  was also observed.
  The energy of  one proton is plotted as a function  of the energy of
  the second  proton. (c) The level and decay schemes  of the relevant nuclei.}
\end{center}
\end{figure}

The  focus of the  first experiment made at the SPIRAL facility
was the characterization of $^{19}$Na through a measurement of  the excitation function  for the
 H$(^{18}$Ne$,p)^{18}$Ne reaction~\cite{Villari2002, Oliveira:2005}.
A beam of $^{18}$Ne$^{4+}$ at 7.2 MeV/u and a  purity of about 80\%
(20  \% $^{18}$O and $<$ 1 \% of $^{18}$F) was purified using   a
stripper foil  at the entrance of the LISE spectrometer. The
resulting  pure beam of $^{18}$Ne$^{10+}$, with an intensity of
5x$10^{5}$ pps and target size of 3.8 x  1.7 mm (FW) was used for
the measurement. A thick  solid hydrogen  cryogenic
target~\cite{Dolegieviez:2006} doubled as a  beam stopper.   The
scattered  protons from  the thick target were detected at forward
angles.  A position sensitive  Si telescope was  used to identify
and measure the energy of  the light charged particles.  The  thick
target made it possible to obtain a  complete and continuous
excitation function over a wide range of energies without changing
the energy of the incoming  beam. The  H($^{18}$O,p)$^{18}$O
reaction was used to calibrate the detectors.  The resulting
$^{19}$Na spectrum is shown in
 Fig.~\ref{fig:na19}(a). Several  broad  peaks
corresponding to new states in $^{19}$Na can be seen.

The  measured  spectrum was  analyzed  using  an R-matrix  formalism to derive the
energies,  widths  and  spins  of  these states~\cite{Oliveira:2005}. A large  Thomas-Erhman shift of 725 keV was
observed for the second excited  state in $^{19}$Na (the first peak in
Fig.~\ref{fig:na19}(a)). This was  explained by  Coulomb effects  in the
proximity of  the proton emission threshold. The  understanding of the
decay  process  was  crucial for interpreting the $^{19}$Na  excitation
function.  The measured two-proton  events (Fig.  \ref{fig:na19}(b))  explain  the presence  of two
prominent peaks between 2  and 4 MeV.  The efficiency for the detection of  two-proton events at forward
angles is enhanced by  kinematic focusing inherent to the inverse kinematics.  The
two-proton events  were interpreted  to arise from  sequential decays.
Three  new states with  large widths were found  in $^{19}$Na
(see figure \ref{fig:na19}(c)). A by-product of this work was the development of a
new experimental  method \cite{Dalouzy:2009} based  on proton-proton
correlations to study inelastically excited states of the  beam.  Excited
states and their decay modes in $^{18}$Ne were also  characterized from the analysis of the  two proton events.

\subsection{States in $^{16}$F}

\begin{figure}
\includegraphics[width=\columnwidth]{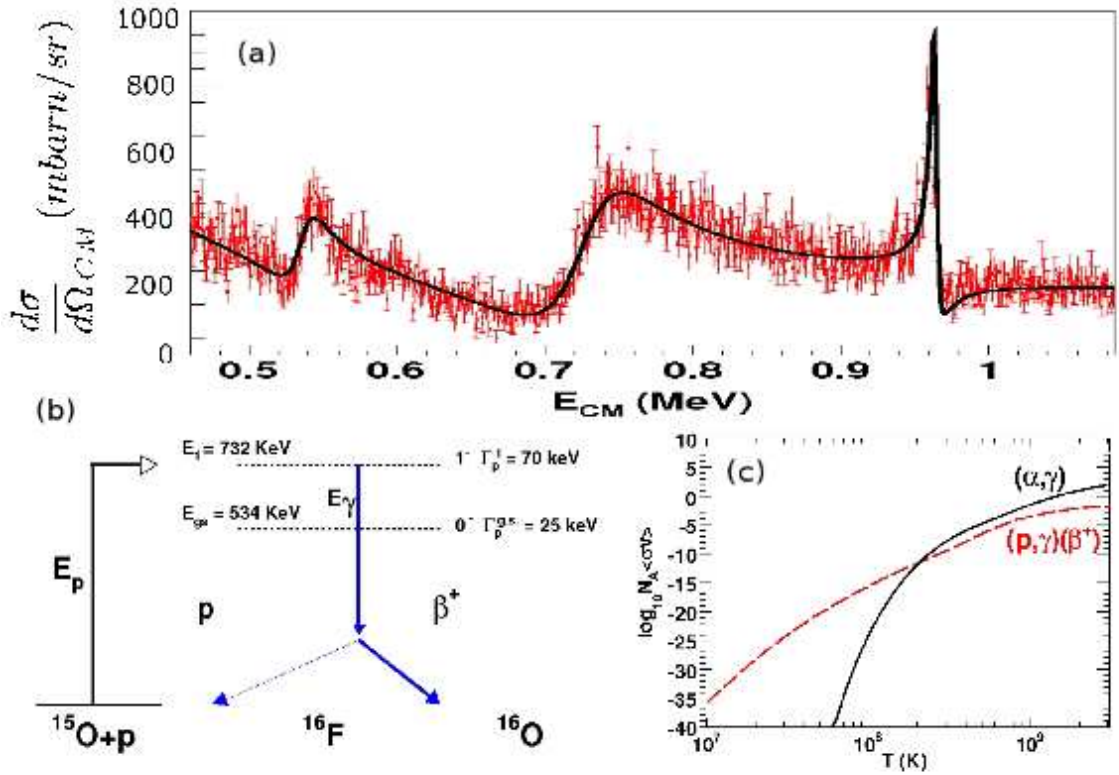}
 \caption{\label{fig:f16} (colour online)    (a)     Excitation    function     for    the
   H($^{15}$O,p)$^{15}$O reaction  measured at 180$^{\circ}$  in the centre of mass.
   The line corresponds to  an R-matrix calculation. (b) Illustrating
   of possible new  process proposed  to bypass  the $^{15}$O
   waiting  point. (c)  Calculated reaction rates  using  the measured
   properties of $^{16}$F.}
\end{figure}

The unbound system $^{16}$F was  studied using  the method described
above for $^{19}$Na~\cite{Stefan:2006,Oliveira:2007}. A beam  of
$^{15}$O at  1.2 MeV/u and an average intensity of 10$^{7}$ pps
bombarded a  21$\mu$m thick polypropylene target. A  beam
contamination of  less than 1 $\%$  of $^{15}$N$^{1+}$ was  achieved
using  a  stripper foil  at  the  entrance  of   the  LISE
spectrometer.   Two stable  beams, $^{14,15}$N  obtained  under
similar experimental conditions  were used for calibration purposes.
Scattered protons  were identified  by their  energy (5  keV
resolution in  the centre  of mass frame) and  time-of-flight.
Figure~\ref{fig:f16}(a) shows the measured excitation function, in
the  range   from  $0.450$  MeV  to  1.1   MeV,  for  the
H($^{15}$O,p)$^{15}$O  reaction where three peaks can be seen. Since
the energy resolution is proportional to
tan($\theta_{lab}$)d$\theta_{lab}$,  protons  were  detected  at
zero degree  with a  restricted  angular acceptance  to  achieve the
best energy resolution.  An energy resolution (FWHM) better than 5
keV in the centre of mass was obtained (as far as we are aware this
is the best energy resolution obtained for charged particles
involving RIB).

An  R-matrix  analysis was  used  to extract the properties  of the
first three states  of $^{16}$F necessary to calculate the
two-proton  capture reaction rates  under  astrophysical conditions.
Two  new   reaction  channels $^{15}$O(p,$\beta^{+}$)$^{16}$O and
$^{15}$O($p$,$\gamma)(\beta^{+}$)$^{16}$O were  proposed to bypass
the $^{15}$O waiting point. Both  reactions eventually proceed
through the $\beta^{+}$-decay  of  the intermediate  unbound ground
state of $^{16}$F (Figure \ref{fig:f16}(b)), which is fed directly
by a proton capture or through  a  proton  capture  to the first
excited  state followed by  $\gamma$ decay. Following the
$\gamma$-emission to the low-energy tail of the ground  state
resonance, the hindrance of the proton decay by the  Coulomb barrier
makes $\beta$   decay a competitive process. The  cross section of
these reactions were calculated by considering the  low energy  tail
of   $^{16}$F  to be quasi-bound~\cite{Stefan:2006,Oliveira:2007}.
The calculated reaction  rates (Fig.~\ref{fig:f16}(c)) indicate that
the  new proposed reactions can compete with   the
$^{15}$O$(\alpha,\gamma)^{19}$Ne breakout reaction.

\subsection{Determination of the $^{46}$Ar (n,$\gamma$) $^{47}$Ar reaction rate}

About  two  decades  ago,  Wasserburg  and   collaborators
identified   correlated  isotopic   anomalies  for   the  neutron-rich
$^{48}$Ca,  $^{50}$Ti  and $^{54}$Cr  isotopes  in certain  refractory
inclusions of the  Allende meteorite~\cite{Lee,Nie}.  For example, the
$^{48}$Ca/$^{46}$Ca ratio  was found to be  250, a factor  of 5 larger
than observed in the  solar system. It was concluded  that these highly unusual
isotopic  compositions were  witness to late-stage nucleosynthesis  processes
which  preceded   the  formation   of  the  solar   nebula.   However,
astrophysical  models   existing  at  that   time  encountered  severe
difficulties  when trying  to reproduce  these observed  anomalies, in
particular those in the  EK-1-4-1 inclusion. A plausible astrophysical
scenario to account for the overabundance of $^{48}$Ca is a weak rapid
neutron-capture  process~\cite{sorl03,Kra}.  In  such  a scenario  the
neutron-rich  stable  $^{46,48}$Ca  isotopes  are  produced  during  a
neutron-capture and $\beta$-decay process from lighter-$Z$ stable seed
nuclei. The main  contribution to the production of  these Ca isotopes
is provided by the $\beta$-decay of their progenitor isobars in the Ar
isotopic chain~\cite{sorl03,Gaud05}. This  hypothesis was derived from
the  fact  that, in  the  $Z<18$  chains,  the measured  $\beta$-decay
lifetimes   of   $^{44}$S  and   $^{45}$Cl   are   shorter  than   the
neutron-capture rates  at the $N=28$ shell  closure. Consequently, the
matter flow in the S and Cl chains is depleted by $\beta$-decay to the
higher $Z$ isotopes  before reaching masses A=46 or  48.  Thus the main
progenitors  of $^{46,48}$Ca are  produced either  directly in  the Ar
chain or  from the $\beta$-decay  of $Z<18$ nuclei  which could subsequently
capture neutrons in the Ar chain. Therefore, the determination of
neutron-capture rates  in the Ar isotopes is  important. Indeed,  a high
neutron-capture  rate   at  A=46  would  reduce  drastically   the  amount  of
progenitors  of $^{46}$Ca  as  the neutron-capture  quickly shifts  the
matter  flow  to  A=48,   enriching  $^{48}$Ca  accordingly.   In  the
following,   the $^{46}$Ar(d,p)$^{47}$Ar   reaction   is  used   as   a
``surrogate"  to determine the  $^{46}$Ar(n,$\gamma$)$^{47}$Ar reaction
rate.
\begin{figure}[t]
\begin{center}
\includegraphics[width=12cm, height=6cm]{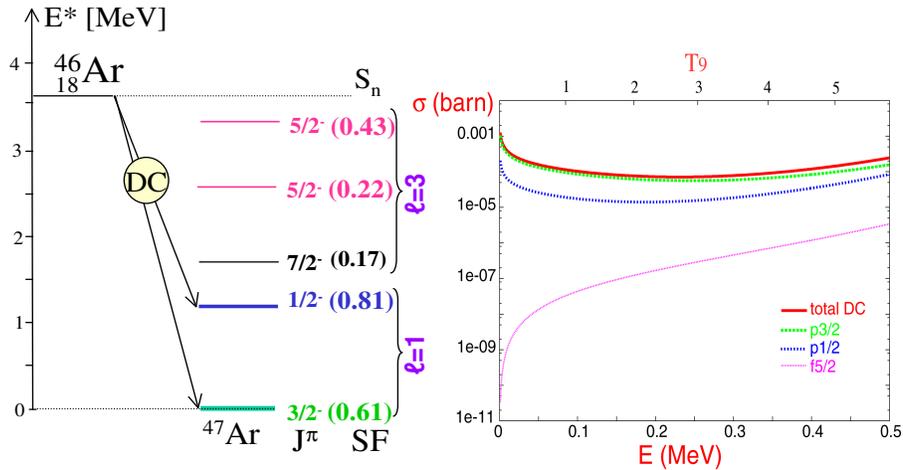}
\end{center}
\caption{ (colour online)  Left: Experimental level scheme of $^{47}$Ar as a function of
  the  excitation energy   E$^*$.  The  main  levels  which   contribute  to  the
  neutron-capture  are  represented   with  their    angular
  momenta $\ell$ and $J$ and  spectroscopic factors (in parenthesis). Right:
  Calculated $\sigma_n$  as a function of the stellar temperature in units of 10$^9$K (top axis) or
as a function of the neutron energy (bottom axis) for the three
levels indicated. The total cross section is shown in red.}
\label{ncapt}
\end{figure}

The $^{46}$Ar nucleus  lies at the N=28 shell  closure, where a
sudden drop of the  neutron-separation energy is  expected  to lead
to a change in the  Q value for the neutron-capture  cross section,
despite the reduction of the shell gap. Consequently, the
$^{46}$Ar(n,$\gamma$)$^{47}$Ar capture cross section ($\sigma_n$)
is, a  priori, expected  to be small as compared  to neighbouring
nuclei. In the  present case of low nuclear  density   and  where
bound  $p$  states   are  present,  the neutron capture  occurs
principally through  direct capture  to bound states (DC). Taking
into account the conservation of spin in the reaction, the neutron
is captured in  $\ell$=1 bound states through the E1  operator
without  a centrifugal barrier,   as the  transferred angular
momentum is $\ell_n$=0 (s-wave  capture). As shown in
Fig.~\ref{ncapt}, the ratio  between an  s-wave (final state  with
$\ell$=1)  and d-wave (final   state   with  $\ell$=3)   direct
neutron capture  rate   is approximately 10$^4$ at a  stellar
temperature of 10$^9$K. This arises from the influence of the
centrifugal barrier which strongly hampers neutron-capture on orbits
with  higher angular momenta.   Therefore, the  contribution of
direct capture to  the g.s.   and first excited  states,  both
having a large  spectroscopic  factors $\ell$=1, dominate. This
gives rise to large neutron capture cross sections for  A=46,
$^{46}$Ar, leading to a small  depletion through  $\beta$-decay at
A=46 as the lifetime of $^{46}$Ar (7.85 s) is longer than the
neutron capture (t$_n$). At A=48, t$_n$ becomes much longer than
t$_\beta$. It follows that neutron captures are halted in the Ar
chain at A=48, accumulating a substantial amount of $^{48}$Ca. These
two features can  explain the high $^{48}$Ca/$^{46}$Ca ratio for a
stellar temperature of 10$^9$K and a neutron density $d_n$ of 6
$\times$10$^{19}$/cm$^3$.

\section{Summary and Perspectives}

A large and varied program, using a range of direct and  compound processes
with re-accelerated beams from SPIRAL in range of 1.2 to 20 MeV/u, to address several important questions in nuclear physics,
has been presented.  Such a program also necessitated the development of a variety of  state-of-the art detectors, such as the  active target
MAYA, the charged particle detectors like MUST/MUST2 and TIARA, the EXOGAM
Ge array and the  versatile high acceptance spectrometer VAMOS.
In the coming  years GANIL concentrate its development and research activities
on  ISOL with the  commissioning of  the SPIRAL2
facility~\cite{SP2}.   The wide  ranging capabilities at  GANIL  will also
continue  to  be  maintained,  also  keeping in  mind  that  the  post
acceleration  of fission fragment beams  at higher  energies could  be one
possible future avenue for the facility.  In  the near future,
beams available at  SPIRAL are  planned to be
extended  \cite{Del09}.    In  particular,   the  use  of   hot  1$^+$
ion-sources,  such as  surface  ionization cavities (already tested)
and FEBIAD sources developed at CERN, are envisaged to enable the
production of  alkali and metallic  beams having melting points  up to
2000$^\circ$K. These, in conjunction with an ECR charge breeder (tested
on-line at CERN ISOLDE), are expected to provide post accelerated beams
with A $\leq$ 100 up to energies of 10 MeV/u.  In this scenario with the
present graphite  targets, neutron  deficient beams such  as $^{38}$Ca
should also be possible with intensities $\leq$10$^4$~pps. In a longer
term perspective,  the use  of target materials having a higher Z should
permit the production of isotopes  of heavier elements ranging from  Sn
up to Ta.

\begin{figure}
\begin{center}
\includegraphics [width=0.8\columnwidth]{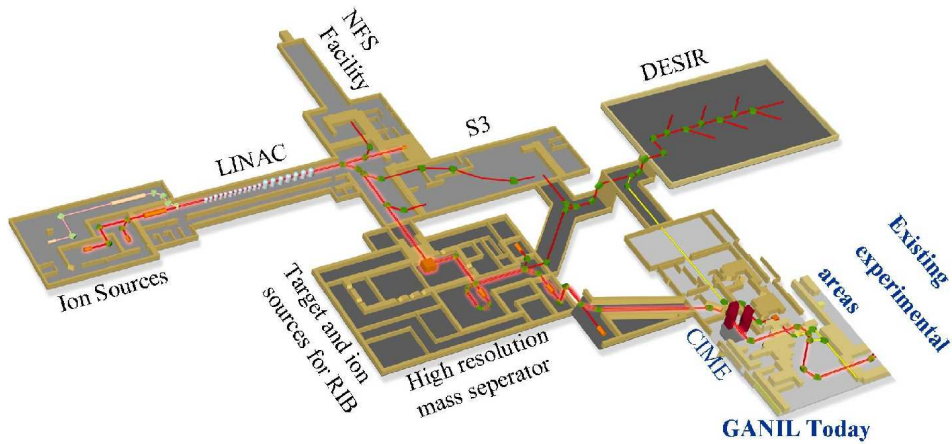}
\caption {(colour online)  The  evolution of the  GANIL facility compared  to that
  shown in Fig. 1, showing the various components of the SPIRAL2 facility.}
\label{spiral2}
\end{center}
\end{figure}

 SPIRAL2  \cite{SP2} is a  major  expansion of  the  SPIRAL facility, and  represents a significant
 element in European ISOL development  program.  The main
 goal of  SPIRAL2 is to  extend the knowledge of  the limit of
 existence and the structure of nuclei far from stability to presently unexplored regions
 of  the nuclear chart,  in particular  in the  medium and  heavy mass
 region (60$\leq$A$\leq$140).   SPIRAL2 is based on a  high power, CW,
 superconducting driver LINAC, delivering up to 5 mA of deuteron beams
 at 40  MeV (200  kW) which will impinge on a C  converter + Uranium  target to
 produce  more  than 10$^{13}$  fissions/s.   The expected
 radioactive ion  beam intensities for  exotic species  in the  mass range
 from A=60  to A=140,  are of the order  of 10$^{6}-10^{10}$ pps.  These   beams will  be available  at
 energies between a few keV/u  (at the dedicated DESIR facility) and 10
 MeV/u (after post acceleration by the CIME cyclotron) in the current experimental halls.  The  same driver will also be used
 to accelerate high intensity  (100$\mu$A to 1 mA), light  and heavy ions
 up to 14.5  MeV/u to produce neutron deficient  and very heavy exotic
 nuclei. In the first  phase the intense  stable beams from
 the LINAC will be used in conjunction with the high rejection
S3 spectrometer~\cite{detectors} to study heavy and super heavy nuclei and to produce a high
flux  of neutrons in the  energy range of  1-40 MeV at  the NFS (Neutrons  for Science)
 facility.  A broad range  of next generation instrumentation, including the
 PARIS gamma-array, the upgrade of EXOGAM electronics (EXOGAM2) and the  FAZIA  and  GASPARD charged particle
 arrays are being planned  to exploit the high intensity radioactive
 ion beams.   Suitable changes to  the existing facility to  account for safety regulations related to the use
 of  these high intensity  beams are in progress. The first intense beams  from the LINAC and
 the re-accelerated beams of fission  fragments are expected  in  2012  and in   2014 respectively.
 The SPIRAL2  project will substantially  increase technical  know-how
of  technical solutions that will  contribute to EURISOL~\cite{EURISOL}
 and  other projects around the world.   With the
 completion  of the  SPIRAL2 project  a  wide diversity  of beams  with
 energies ranging from a few keV  to 95 MeV/u coupled to broad array  of detection systems  will be available  at GANIL. The
 entire accelerator complex  will be able to provide up  to five stable
 and/or radioactive ion beams simultaneously.  Such developments
hold  promise  in advancing our  understanding of the  origin of simple
 patterns in complex nuclei.

\ack We would  like to acknowledge the help of all our colleagues  at GANIL and elsewhere
in  France and Europe in preparing this article.

\end{document}